\documentclass[aps,onecolumn]{revtex4}
\usepackage{graphicx}
\usepackage{epsfig}
\usepackage{epstopdf}
\usepackage{amsfonts}
\usepackage{amssymb}
\usepackage{amsbsy}
\usepackage{amsmath}
\usepackage{mathrsfs}
\usepackage{latexsym}
\usepackage{natbib}
\usepackage{bm}
\usepackage{color}
\usepackage[a4paper, total={6.5in, 9.7in}]{geometry}
\DeclareMathOperator{\sech}{sech}

\DeclareMathOperator{\taninv}{arctan}
\usepackage{braket}
\usepackage{slashed}
\usepackage{pgfplots}
\numberwithin{equation}{section}

\usepackage{tikz}
\usetikzlibrary{shapes,arrows,shadows}
\begin{document}

\title{Low-energy effective quantum field theoretic description of excitations about soliton configurations}

\author{Susobhan Mandal}
\email{sm17rs045@iiserkol.ac.in}

\affiliation{ Department of Physical Sciences,\\ 
Indian Institute of Science Education and Research Kolkata,\\
Mohanpur - 741 246, WB, India }

%\date{\today}

\begin{abstract}
\begin{center}
\underline{\textbf{Abstract}}
\end{center}
Solitons are the classical field configurations connecting two trivial vacua. These are also the solutions of classical field equations of motion with particle-like properties. Moreover, they are localized in space, having finite energy, and are stable against decay into radiation. The coherent state description of kink-solitons is discussed in the present article. Further, the relation between topological solitons and occupation numbers corresponding to low momentum excitations are also discussed coherently. The description of the low energy excitations about solitons in quantum field theory is the main theme of this article. Further, a few physical observables, namely some low order correlation functions, are computed up to certain integral forms. Furthermore, we have shown that it is possible to detect the presence of soliton-like classical configurations in many-particle systems from the nature of the one-point function and non-conservation of momentum feature of one-point, two-point and three-point functions in this low-energy effective field theory of these excitations.    
\end{abstract}

\maketitle

\section{Introduction}
The present article deals with the important theoretical framework that allows us to replace classical solutions with a quantum system. The idea is to consider the classical solutions as the expectation values of the field operators \cite{Nair:2005iw} \textit{w.r.t} the coherent states, made out of particle number eigenstates. We first present a brief review of the coherent state approach to soliton physics, and then we mention the quanta, responsible for the topological charge and for the energy of soliton configurations \cite{Rajaraman:1982is}. This is our first result, presented in section (\ref{section1}). 

Coherent states \cite{nozari2005quantum, fujii2001introduction, perelomov1977generalized, sivakumar2000studies} are first discovered in the context of quantum-mechanical simple harmonic oscillator since they are the states with minimized uncertainty. Hence, they are in a way closest to being classical since the expectation values of canonically conjugate operators \textit{w.r.t} these states follow classical dynamics. In quantum field theory, the coherent states are defined to be the eigenstates of the annihilation operators and \textit{w.r.t} this basis, the functional integral formulation of field theories can be constructed. That is why we choose to work with these states.

With some simple examples of kink-solitons, the quantum-mechanical explanation of conservation of topological charge, encoded in an infinite occupation of long-wavelength quanta is demonstrated. In order to make this presentation self-contained, a comparison between a topological kink-soliton with a non-topological soliton with zero topological charge is presented. On the other hand, it can also be shown that the finiteness of the occupation number leads to instability of the classical solution since it decays to the trivial vacuum configuration. 

Considering the normal modes about a kink-soliton configuration, the effective low-energy description of excitations about that soliton configuration is important in order to understand the behaviour in a large distance scale, which is presented in section (\ref{section3}). The knowledge of low-energy behaviour essentially gives us information about the correlation functions of these excitations at large distance scales without worrying about short-distance behavior. The interactions between excitations about soliton configurations at a low-energy scale are important in many low-energy processes that take place at the level of nucleons \cite{kalafatis1992soliton}, in Ising models \cite{kehrein2002soliton}, in spin chains \cite{mukhopadhyay2015rogue}, etc.

Solitons are non-perturbative classical configurations that connect the trivial vacua in a non-trivial manner \cite{mandal2018characteristics}. They can be found by minimizing the energy of a classical configuration. Moreover, the soliton configurations have non-trivial energy density \cite{mandal2018characteristics}. Hence, it is expected that excitations about these configurations carry long-range properties of the system in terms of correlations between asymptotic (at spatial infinity) classical vacua, which are presented here. Further, it is shown in section (\ref{section4}) that unlike the interacting quantum field theories (QFT) \textit{w.r.t} the trivial vacuum where excitations at the quantum level (in loop corrections) follow 4-momentum conservation, here such conservation is violated for these excitations at the level of quantum corrections. For the sake of mathematical simplicity, we restrict ourselves to scalar solitons throughout the discussion. 

\section{Coherent state description}\label{section1}
In this section, we briefly review the formulation of the coherent state description of soliton configurations. We provide one example each for both the non-topological and topological solitons in section (\ref{section1.2}) and section (\ref{section1.3}). In these two sections, the role of soft modes is also highlighted. On the other hand, in section (\ref{section1.4}), our first result, namely the residue of the pole at $k=0$ in a topological soliton configuration is proportional to the topological charge of that configuration is presented.

\subsection{Coherent states}
A generic classical solution $\phi_{c}(x)$ in a given scalar quantum field theory in $d+1$-dimensional spacetime can always be expressed in terms of Fourier modes in the following manner
\begin{equation}\label{1}
\phi_{c}(x)=\int\frac{d^{d}k}{\sqrt{(2\pi)^{d}\omega_{k}}}\Big[\alpha_{k}(t)e^{-ik.x}+\alpha_{k}^{*}(t)e^{ik.x}\Big],
\end{equation}
where $\{\alpha_{k}(t)\}$s are the time-dependent classical expansion coefficients, $k$ is a $d+1$-vector with $k^{0}=\omega_{k}$ being the dispersion relation whose form can be found out from the equation of motion without interaction. However, for a static classical solution, $\{\alpha_{k}(t)\}$s become time-independent, and $\omega_{\vec{k}}$ becomes zero. The metric convention in Natural units is chosen as $(+,-,-,\ldots,-)$. During quantization, $\{\alpha_{k}\}, \ \{\alpha_{k}^{*}\}$s are replaced by annihilation operators $\{\hat{a}_{k}\}$ and creation operators $\{\hat{a}_{k}^{\dagger}\}$, respectively. The above-mentioned operators follow the following commutation algebra
\begin{equation}
\begin{split}
[\hat{a}_{k},\hat{a}_{p}]=0 & =[\hat{a}_{k}^{\dagger},\hat{a}_{p}^{\dagger}]\\
[\hat{a}_{k},\hat{a}_{p}^{\dagger}] & =\delta^{d}(k-p).
\end{split}
\end{equation}
Coherent states by definition are the eigenstates of annihilation operators
\begin{equation}
\hat{a}_{k}\ket{\text{coh}}=\alpha_{k}\ket{\text{coh}}.
\end{equation}
However, the values of $\{\alpha_{k}\}$s corresponding to coherent states associated with a soliton are not fixed yet. From our previous discussion, it must be cleared that we have to construct the coherent state in such a way that the expectation value of field operator \textit{w.r.t} this state mimics the classical solution. This can only be done in the following way
\begin{equation}
\ket{\text{Coh}}=\prod_{k}e^{-\frac{N_{k}}{2}}\Big[\sum_{n_{k}}\frac{N_{k}^{\frac{n_{k}}{2}}}{\sqrt{n_{k}!}}\ket{n_{k}}\Big],
\end{equation}
where $\alpha_{k}=\sqrt{N_{k}}$ and $\hat{a}_{k}^{\dagger}\hat{a}_{k}\ket{n_{k}}=n_{k}\ket{n_{k}}$. $\{\ket{n_{k}}\}$ are the eigenstates of the number operator of quanta with momentum $k$.

\subsection{Coherent state picture of Non-Topological solitons}\label{section1.2}
In order to clearly distinguish the role of topology, we discuss non-topological solitons first and then topological solitons. Let us consider a $1+1$-dimension classical field theory with the following Lagrangian density
\begin{equation}
\mathcal{L}=(\partial\phi)^{2}-m^{2}\phi^{2}+g^{2}\phi^{4},
\end{equation} 
where $m^{2},g^{2}>0$. This theory has stable vacuum at $\phi(x)=0$, however, it becomes unstable for large field values. There is also a solution which interpolates $\phi=0$ and $\phi=\frac{m}{g}$. A static solution of the equations of motion is given by the following
\begin{equation}
\phi_{sol}(x)=\frac{m}{g}\sech(mx),
\end{equation}
which shows that at $x=\pm\infty, \ \phi=0$ and at $x=0, \ \phi=\frac{m}{g}$. The energy of the above configuration is 
\begin{equation}
E_{non-top}=\frac{2m^{3}}{3g^{2}}.
\end{equation}
As we have described earlier in (\ref{1}), we can write the above solution as
\begin{equation}
\phi_{sol}(x)=\sqrt{R}\int\frac{dk}{\sqrt{4\pi|k|}}\Big[\alpha_{k}e^{ikx}+\alpha_{k}^{*}e^{-ikx}\Big],
\end{equation}
where $R$ is the regularized volume of the space. In the present context of $1+1$-dimensional field theory, $R=2\pi L$ where $L$ is the length of the system considered here. $\{\alpha_{k}\}$s are chosen such that they are the Fourier transform of the classical solution 
\begin{equation}
\begin{split}
\alpha_{k} & =\frac{1}{\pi}\sqrt{\frac{|k|}{R}}\int_{-\infty}^{\infty}\phi_{sol}(x)e^{-ikx}dx=\frac{1}{\pi}\sqrt{\frac{|k|}{R}}\frac{m}{g}\int_{-\infty}^{\infty}\frac{2}{e^{2mx}+1}e^{(ik+m)x}dx\\
 & =-\sqrt{\frac{|k|}{R}}\frac{2i}{g}\sum_{n=0}^{\infty}e^{-\frac{k}{2m}(2n+1)\pi}e^{(n+\frac{1}{2})i\pi}=\sqrt{\frac{|k|}{R}}\frac{2}{g}\frac{e^{-\frac{k}{2m}\pi}}{1+e^{-\frac{k}{m}\pi}}=\sqrt{\frac{|k|}{R}}\frac{1}{g}\sech\left(\frac{k}{2m}\pi\right).
\end{split}
\end{equation}
Therefore, we obtain the following relation
\begin{equation}
\alpha_{k}^{*}\alpha_{k}=\frac{|k|}{R}\frac{1}{g^{2}}\sech^{2}\left(\frac{k}{2m}\pi\right),
\end{equation}
Using the above the information, we can express the energy as 
\begin{equation}
E_{non-top}=R\int_{-\infty}^{\infty}dk\frac{k^{2}}{2R}\frac{1}{g^{2}}\sech^{2}\left(\frac{k}{2m}\pi\right)=\frac{2m^{3}}{g^{2}}.
\end{equation}
Hence, it follows from the above result that we can write soliton state in the following way
\begin{equation}
\ket{sol}=\prod_{k}\otimes\ket{\alpha_{k}},
\end{equation}
where
\begin{equation}
\begin{split}
\ket{\alpha_{k}} & =e^{-\frac{1}{2}|\alpha_{k}|^{2}}e^{\alpha_{k}\hat{a}_{k}^{\dagger}}\ket{0}=e^{-\frac{1}{2}|\alpha_{k}|^{2}}\sum_{n_{k}=0}^{\infty}\frac{\alpha_{k}^{n_{k}}}{\sqrt{n_{k}!}}\ket{n_{k}},
\end{split}
\end{equation}
and $\ket{n_{k}}$s are the number eigenstates of quanta with momentum $k$. Now we define particle number operator $\hat{N}=\sum_{k}\hat{N}_{k}, \ \hat{N}_{k}=\hat{a}_{k}^{\dagger}\hat{a}_{k}$. We also obtain the relation $N_{k}\equiv\bra{\alpha_{k}}\hat{N}_{k}\ket{\alpha_{k}}=\alpha_{k}^{*}\alpha_{k}$. Hence, the total number of particles in the soliton state is 
\begin{equation}
N\equiv\int_{k}N_{k}=\int_{k}|\alpha_{k}|^{2}=\frac{m^{2}}{g^{2}}\left(\frac{8\ln(2)}{2\pi}\right), \ \int_{k}=R\int_{-\infty}^{\infty}dk.
\end{equation}
It is important here to emphasize that the dominant contribution for both energy and total particle number comes from the quanta of momentum $k\leq m$. That is the reason why the size of the soliton energy density is $m^{-1}$. The finite number of quanta contained in this state reflects the fact that quantum mechanically vacuum state $\phi=0$ is unstable
\begin{equation}\label{2}
\braket{0|sol}=e^{-\frac{1}{2}\int_{k}|\alpha_{k}|^{2}}=e^{-\frac{N}{2}}\approx e^{-\frac{m^{2}}{\hbar g^{2}}}.
\end{equation}
From this point of view, we can say that creation of non-topological soliton reflects the instability of the vacuum $\phi=0$. Further, in classical limit only it becomes the energy eigenstate, and amplitude in (\ref{2}) vanishes. 

\subsection{Coherent state picture of Topological soliton}\label{section1.3}
Next, we consider a topological soliton described by
\begin{equation}
\mathcal{L}=(\partial\phi)^{2}-g^{2}(\phi^{2}-\frac{m^{2}}{g^{2}})^{2},
\end{equation}
and the reason behind calling this configuration topological is mentioned in \cite{Rajaraman:1982is, vachaspati2006kinks}. There exist solutions of the classical equation of motion that describe configurations connecting two vacuum configurations $\phi=\pm\frac{m}{g}$. One such solution is given by the following
\begin{equation}
\phi_{c}(x)=\pm\frac{m}{g}\tanh(mx),
\end{equation}
where $+,-$ signs refer to kink and anti-kink solutions, respectively. If we follow the previous prescription, we find that
\begin{equation}
\begin{split}
\alpha_{k} & =\frac{m}{g}\sqrt{\frac{|k|}{R}}\int_{-\infty}^{\infty}\tanh(mx)e^{-ikx}dx=\frac{m}{g}\sqrt{\frac{|k|}{R}}\int_{-\infty}^{\infty}\frac{1-e^{-2mx}}{1+e^{2mx}}e^{-ikx}dx\\
 & =-\frac{2\pi i}{g}\sum_{n=0}^{\infty}e^{-\frac{k}{m}(n+\frac{1}{2})\pi}=-\frac{\pi  i}{g}\text{cosech}\left(\frac{\pi k}{2m}\right).
\end{split}
\end{equation}
Hence, the energy of the topological configuration is given by
\begin{equation}\label{3}
E_{top}=\int_{k}|k|N_{k}=\frac{8m^{3}}{3g^{2}}.
\end{equation}
Unlike the non-topological case, here we have a first-order pole at $k=0$ in $N_{k}$ which shows $N_{k}$ exhibits $\frac{1}{k}$ singularity for lower momentum values. This singularity has its own importance in terms of describing conserved topological charge in quantum mechanics. As a result of such a pole in the IR region, the number of particles diverges logarithmically
\begin{equation}
N=\int_{0}^{k_{0}}dk \ N_{k}\approx\frac{2\pi}{g^{2}}\int_{0}^{k_{0}}dk\frac{k}{\left(\frac{k\pi}
{m}\right)^{2}}\simeq \frac{2m^{2}}{\pi g^{2}}\int_{0}^{k_{0}}\frac{dk}{k}\simeq \ln(\infty).
\end{equation} 
The quanta $k\rightarrow0$ which contribute to the above divergence do not contribute to the energy expression for the configuration since the presence of an extra $|k|$ factor in the integrand makes the pole disappeared in (\ref{3}). Unlike the non-topological case, in this case, creating soliton out of the vacuum is impossible in quantum theory since
\begin{equation}
\braket{0|sol}=e^{-\frac{1}{\hbar}\infty}=0,
\end{equation} 
and similarly all the amplitudes between a finite number of particle states and $\ket{sol}$ vanish. This is because even with a finite $\hbar$ in quantum theory, soliton has an infinite occupation number of zero momentum mode. That is why soliton is completely a non-perturbative effect in quantum field theory.

Classically the topological charge is coming from the difference of values of the field at the boundaries or in other words, it depends on the asymptotic behavior of fields. On the other hand, in the quantum description, this charge is determined by the infinite wavelength or the zero momentum quanta (soft modes) since topology is about the global properties of the configuration which means it emerges from long-range behavior, captured by the long-wavelength quanta. Because of these distinguishing properties of quanta, contributing to the energy of the soliton configuration and carrying the information of the topological charge, authors in \cite{dvali2015towards, grunding2017towards, book:864295} talk about decomposing a coherent state into a tensor product state in which one part carries information about the topology and another part carries information about the energy of the soliton configuration. However, we do not go into those details in the present article. Instead, we want to make a comment on the relation between the topological charge and the residue of the pole of $N_{k}$ at $k=0$ momentum mode.

\subsection{Topological charge and the residue of the pole of $N_{k}$}\label{section1.4}   
For this case, we need a soliton configuration which can attain one of the many possibilities of vacuum. One such example is the Sine-Gordon model, described by the following Lagrangian density
\begin{equation}
\mathcal{L}=\frac{1}{2}(\partial\phi)^{2}-\frac{m^{2}}{\beta^{2}}(1-\cos\beta\phi).
\end{equation}
For the sake of simplicity in calculation, we take $m=\beta=1$. The solution of the equation of motion is described by the following kink-soliton
\begin{equation}
\phi(x)=4\arctan(e^{x}),
\end{equation}
where the system has an infinite number of vacuum at $\phi=2n\pi$ with $n$ being an integer ($n\in\mathbb{Z}$). However, in order to have more than unit charge, we need to consider a multi-kink configuration, constructed using the additive ansatz described in \cite{Mandal:2018pqo}.
\begin{figure}
\begin{align}
\includegraphics[height=5.5cm,width=7.5cm]{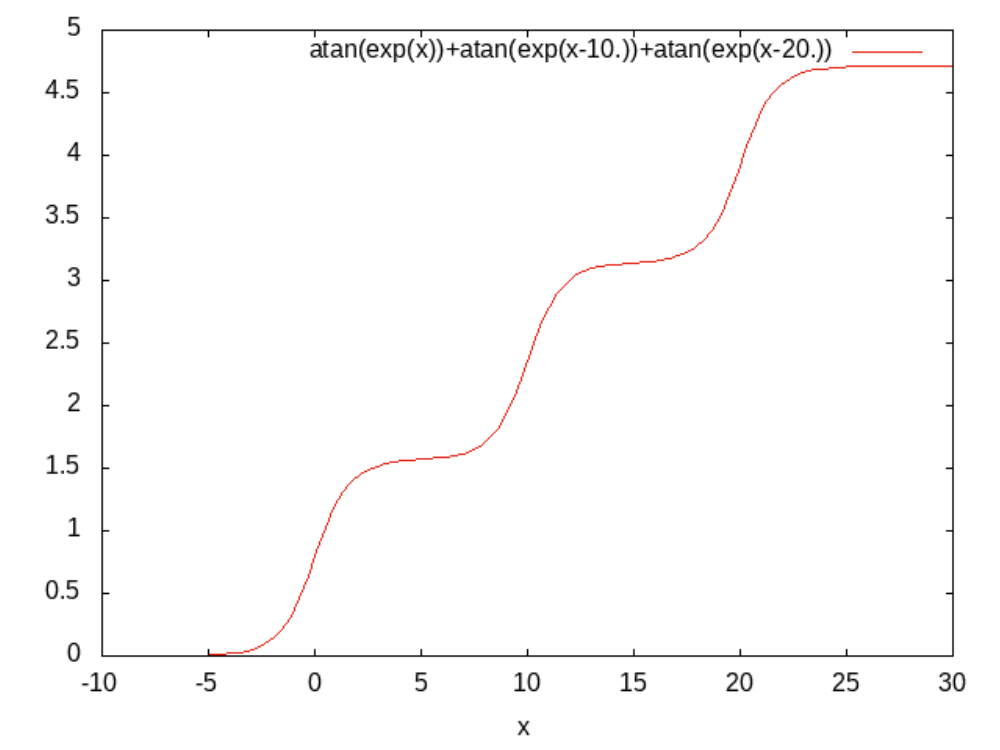}
\end{align}
\caption{An example of a multi-kink soliton configuration with topological charge $N=3$.}
\begin{align}
\includegraphics[height=5.5cm,width=7.5cm]{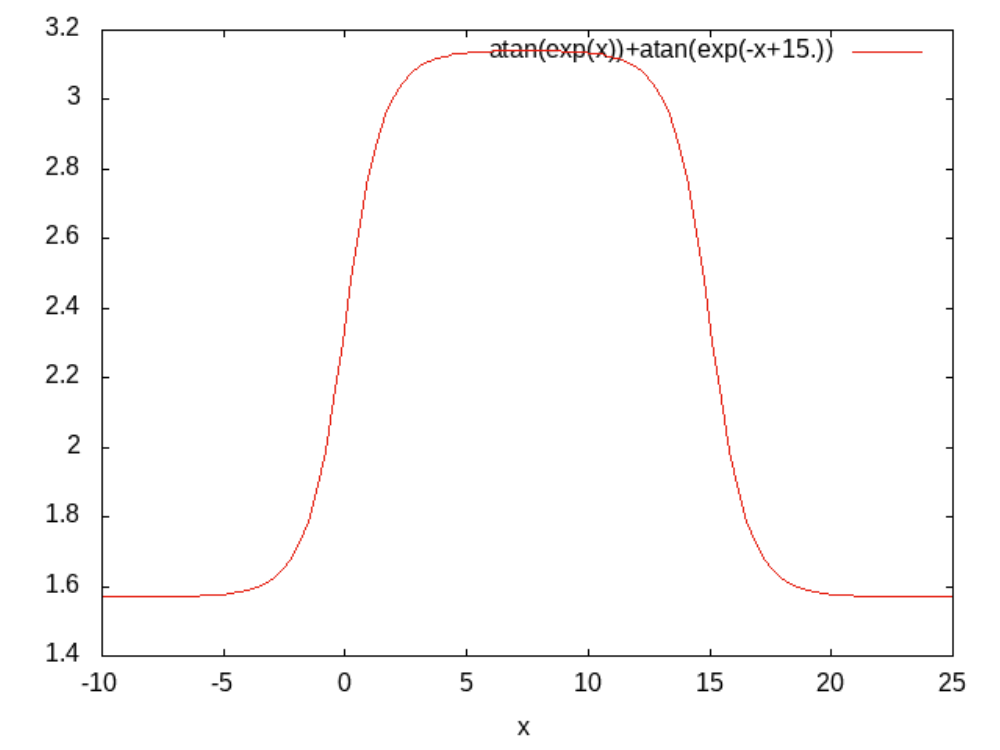}
\end{align}
\caption{An example of a soliton configuration with zero topological charge, neighbouring kink, and anti-kink.}
\end{figure}
We choose the following configuration
\begin{equation}\label{4}
\phi_{c}(x)=4\arctan e^{x}+4\arctan e^{x-a}+4\arctan e^{x-2a}+\ldots+4\arctan e^{x-(N-1)a},
\end{equation}
where the charge $\mathcal{Q}_{\text{top}}\equiv\frac{1}{2\pi}\int_{-\infty}^{\infty}\frac{d\phi_{c}}{dx}dx=N$ and $a$ is the distance between neighbouring kinks. Two such configurations are shown in FIG. 1 and FIG. 2 up to an overall factor $4$. Using the following property of the Fourier transform 
\begin{equation}
\begin{split}
\mathcal{G}(k) & =\int_{-\infty}^{\infty}\mathcal{G}(x)e^{-ikx}dx\\
\implies\int_{-\infty}^{\infty} & \mathcal{G}(x-a)e^{-ikx}dx=\int_{-\infty}^{\infty}\mathcal{G}(x)e^{-ikx}e^{-ika}dx=\mathcal{G}(k)e^{-ika},
\end{split}
\end{equation} 
the Fourier transformation of $\phi_{c}(x)$ in (\ref{4}) can be expressed as 
\begin{equation}
(\text{fourier transform of} \ 4\arctan e^{x})\times(1+e^{-ika}+e^{-2ika}+\ldots+e^{-ik(N-1)a}),
\end{equation}
and the Fourier transform of $\arctan e^{x}$ is given by 
\begin{equation}
\begin{split}
\mathcal{I}_{\lambda} & \equiv\int_{-\infty}^{\infty}(\arctan e^{\lambda x})e^{-ikx}dx=\frac{i\pi}{k}\frac{e^{-\frac{\pi k}{2\lambda}}}{1+\frac{\pi k}{\lambda}}\\
\implies\mathcal{I} & =\int_{-\infty}^{\infty}(\arctan e^{x})e^{-ikx}dx=\mathcal{I}_{\lambda=1}.
\end{split}
\end{equation}
Hence, we obtain the following relations
\begin{equation}
\begin{split}
\implies\alpha_{k}=\sqrt{k}\mathcal{I} & (1+e^{-ika}+e^{-2ika}+\ldots+e^{-ik(N-1)a})=\frac{i\pi}{2\sqrt{k}}\sech\left(\frac{\pi k}{2\lambda}\right)\frac{1-e^{-ikNa}}{1-e^{-ika}}\\
\implies N_{k} & \propto\frac{\pi^{2}}{k}\sech^{2}\left(\frac{\pi k}{2\lambda}\right)\Big|\frac{1-e^{-ikNa}}{1-e^{-ika}}\Big|^{2},
\end{split}
\end{equation}
which shows that the residue of the pole of $N_{k}$ is given by
\begin{equation}
\lim_{k\rightarrow0}N_{k}\rightarrow\frac{\pi^{2}}{k}N^{2}\implies\text{Res}\{N_{k}\}=\pi^{2}N^{2},
\end{equation}
or in other words, $\lim_{k\rightarrow0}N_{k}$ is proportional to $N^{2}$, square of the topological charge. Similarly, if we had considered kink and anti-kink in the nearest neighbourhood, then the topological charge would be zero since $(1-e^{-ika})(1-e^{ika})=2(1-\cos(ka))$ multiplied with $N_{k}$ makes the pole at $k=0$ removed. Therefore, we have proven that the residue of the pole of $N_{k}$ at $k=0$ in a topological soliton configuration is proportional to the square of the topological charge of that configuration.

\section{Normal modes about kink soliton} 
In this section, we briefly review the normal modes about the kink-soliton in $\phi^{4}$ theory. Once we get the normal modes, the fluctuations about the solitons in terms of the normal modes can be expressed explicitly. These modes are nothing but the solution of the following eigenvalue equation
\begin{equation}\label{5}
\begin{split}
\Big[-\frac{d^{2}}{dx^{2}}+U(x)\Big] & f_{j}(x)=\omega_{j}^{2}f_{j}(x), \ U(x) =V''(\phi_{k}(x))=3\lambda\phi_{K}^{2}-m^{2}=m^{2}\Big[2-\frac{3}{\cosh^{2}[\frac{mx}{\sqrt{2}}]}\Big],
\end{split}
\end{equation}
where $\phi_{K}$ is the kink solution. Since $U(\pm\infty)=2m^{2}$, we have continuum spectrum of eigenvalues. However, there could also be discrete spectrum. It is important here to note that zero is an eigenvalue since the solution $\phi_{K}$ satisfies the following relation
\begin{equation}
-\frac{d^{2}}{dx^{2}}\phi_{K}+V'(\phi_{K})=0.
\end{equation}
Taking one more derivative \textit{w.r.t} $x$ gives the following equation
\begin{equation}
-\frac{d^{2}}{dx^{2}}\left(\frac{d\phi_{K}}{dx}\right)+V''(\phi_{k})\frac{d\phi_{k}}{dx}=0,
\end{equation}
which means $f_{0}(x)=\frac{d\phi_{k}}{dx}$ is the eigenfunction corresponding to eigenvalue $\omega_{0}=0$. This zero eigenvalue corresponds to the spatial translational invariance of the system since an infinitesimal global translation of the kink configuration does not cost any energy.

In order to go further, we have to solve the Schrodinger like differential equation (\ref{5}). Fortunately, the eigenvalues and eigenfunctions of the above differential operator have already been found explicitly, mentioned in \cite{vachaspati2006kinks, book:864295}. Further, there is one more discrete eigenvalue which is $\omega_{1}^{2}=\frac{3m^{2}}{2}$ with corresponding eigenfunction $f_{1}(x)=\frac{\sinh(\frac{mx}{\sqrt{2}})}{\cosh^{2}(\frac{mx}{\sqrt{2}})}$. The continuum modes are again labelled by momentum $k$ and given by
\begin{equation}
f_{k}(x)=e^{ikx}\Big[3m^{2}\tanh^{2}\left(\frac{mx}{\sqrt{2}}\right)-m^{2}-2k^{2}-3\sqrt{2}imk\tanh\left(\frac{mx}{\sqrt{2}}\right)\Big],
\end{equation}
with eigenvalue $\omega_{k}=\sqrt{k^{2}+2m^{2}}$. In the asymptotic limit $x\rightarrow\infty$, the above solutions tend to become the following
\begin{equation}
f_{k}(x)=\sqrt{4(m^{2}-k^{2})^{2}+18m^{2}k^{2}}e^{i(kx-\delta(k))},
\end{equation}
where the phase shift is given by
\begin{equation}
\delta(k)=\arctan\left(\frac{3\sqrt{2}}{2}\frac{mk}{m^{2}-k^{2}}\right).
\end{equation}
Note that $\delta(k)=-\delta(-k).$

\section{Low-energy description of fluctuations around solitons}\label{section3}

\subsection{Wilson prescription for describing low energy action}
Here, we briefly review the low-energy effective field theory using Wilson's prescription. Suppose we are given an action $S[\phi]$ defined at a particular scale momentum $\Lambda$ where the low- and high-energy degrees of freedom are interacting via the interaction term in the Lagrangian density. Wilson found a way to integrating out the fast-modes or high-energy degrees of freedom and was able to write down an effective action for low energy degrees of freedom, describing theory at large length-scale which we are often interested in. This prescription is known as Wilsonian Renormalization \cite{sonoda2006wilson, wilson1975renormalization, epelbaum2017wilsonian, valderrama2016power}.

Hence, we can separate out the low energy and high energy degrees of freedom from the  scalar field in the following way
\begin{equation}
\Phi(x)=\underbrace{\int_{|k|<\frac{\Lambda}{\delta}}\frac{d^{d}k}{(2\pi)^{d}}\phi(k)e^{-ik.x}}_{\phi_{<}(x)}+\underbrace{\int_{\frac{\Lambda}{\delta}<|k|<\Lambda}\frac{d^{d}k}{(2\pi)^{d}}\phi(k)e^{-ik.x}}_{\phi_{>}(x)}, \ \delta>1.
\end{equation} 
In general, an action contains two parts of which one is the quadratic part, known as free theory and the higher-order or interacting part which we call interacting field theory
\begin{equation}
S[\Phi]=S_{0}[\Phi]+S_{I}[\Phi],
\end{equation}
where $S_{0},S_{I}$ are the free and interacting parts of the action, respectively. In free field theory, it is easy to separate out the low- and high-energy degrees of freedom easily which is not the case for the interacting part, therefore, we can write
\begin{equation}
S[\Phi(x)]=S_{0}[\phi_{<}]+S_{0}[\phi_{>}]+S_{I}[\phi_{<};\phi_{>}].
\end{equation}
Using the above information, we can write the partition function as follows
\begin{equation}
\mathcal{Z}=\int\mathcal{D}\Phi \ e^{iS[\Phi]}=\int\mathcal{D}\phi_{<}\mathcal{D}\phi_{>}e^{iS_{0}[\phi_{<}]+iS_{0}[\phi_{>}]+iS_{I}[\phi_{<};\phi_{>}]}.
\end{equation}
Now we rearrange the terms in the following manner
\begin{equation}
\begin{split}
\mathcal{Z} & =\int\mathcal{D}\phi_{<} \ e^{iS_{0}[\phi_{<}]}\Big[\int\mathcal{D}\phi_{>} e^{iS_{0}[\phi_{>}]+iS_{I}[\phi_{<};\phi_{>}]}\Big]\\
 & =\int\mathcal{D}\phi_{<} \ e^{iS_{0}[\phi_{<}]}\underbrace{\int\mathcal{D}\phi_{>} \ e^{iS_{0}[\phi_{>}]}}_{\text{constant}}\Bigg[\frac{\int\mathcal{D}\phi_{>} e^{iS_{0}
[\phi_{>}]+iS_{I}[\phi_{<};\phi_{>}]}}{\int\mathcal{D}\phi_{>} \ e^{iS_{0}[\phi_{>}]}}\Bigg]\\
 & =\text{constant}\times\int\mathcal{D}\phi_{<} \ e^{iS_{0}[\phi_{<}]+iS'[\phi_{<}]}\\
e^{iS'[\phi_{<}]} & =\Bigg[\frac{\int\mathcal{D}\phi_{>} e^{iS_{0}
[\phi_{>}]+iS_{I}[\phi_{<};\phi_{>}]}}{\int\mathcal{D}\phi_{>} \ e^{iS_{0}[\phi_{>}]}}\Bigg].  
\end{split}
\end{equation}
Apart from the trivial constant factor which shifts the effective action by constant amount, we can write
\begin{equation}\label{5.1}
\begin{split}
\mathcal{Z} & =\int\mathcal{D}\phi_{<} \ e^{iS_{\text{eff}}^{(\text{low})}[\phi_{<}]},  \ S_{\text{eff}}^{(\text{low})}[\phi_{<}]=S_{0}[\phi_{<}]+S'[\phi_{<}]\\
S'[\phi_{<}] & =-i\ln\Bigg[\frac{\int\mathcal{D}\phi_{>} e^{iS_{0}
[\phi_{>}]+iS_{I}[\phi_{<};\phi_{>}]}}{\int\mathcal{D}\phi_{>} \ e^{iS_{0}[\phi_{>}]}}\Bigg],
\end{split}
\end{equation}
and if the interaction is perturbative in nature, then we can expand the logarithm with a series expansion
\begin{equation}\label{6}
\begin{split}
S'[\phi_{<}] & =-i\ln\Bigg[1+i\frac{\int\mathcal{D}\phi_{>} \ e^{iS_{0}[\phi_{>}]}S_{I}[\phi_{<};\phi_{>}]}{\int\mathcal{D}\phi_{>} \ e^{iS_{0}[\phi_{>}]}}-\frac{1}{2!}\frac{\int\mathcal{D}\phi_{>} \ e^{iS_{0}[\phi_{>}]}S_{I}^{2}[\phi_{<};\phi_{>}]}
{\int\mathcal{D}\phi_{>} \ e^{iS_{0}[\phi_{>}]}}+\ldots\Bigg]\\
 & =\frac{\int\mathcal{D}\phi_{>} \ e^{iS_{0}[\phi_{>}]}S_{I}[\phi_{<};\phi_{>}]}{\int\mathcal{D}\phi_{>} \ e^{iS_{0}[\phi_{>}]}}-\frac{i}{2!}\Bigg[\frac{\int\mathcal{D}\phi_{>} \ e^{iS_{0}[\phi_{>}]}S_{I}[\phi_{<};\phi_{>}]}{\int\mathcal{D}\phi_{>} \ e^{iS_{0}[\phi_{>}]}}\Bigg]^{2}\\
 & +\frac{i}{2!}\frac{\int\mathcal{D}\phi_{>} \ e^{iS_{0}[\phi_{>}]}S_{I}^{2}[\phi_{<};\phi_{>}]}{\int\mathcal{D}\phi_{>} \ e^{iS_{0}[\phi_{>}]}}+\ldots,
\end{split}
\end{equation} 
where $S_{\text{eff}}^{(\text{low})}[\phi_{<}]$ is the low-energy effective action 
\cite{avramidi2009low, golterman2016low}.

\subsection{Low energy effective action}
In this section, we derive the quadratic part of the low-energy effective action, our second result. In our case, we can write the low- and high-energy modes as follows
\begin{equation}
\begin{split}
\phi_{<}(x) & =\int_{0}^{\Lambda_{0}}\frac{d^{2}k}{(2\pi)}\varphi(k)f_{k}(x)e^{-ik^{0}x}\\
\phi_{>}(x) & =\int_{\Lambda_{0}}^{\Lambda_{\text{UV}}}\frac{d^{2}k}{(2\pi)}\varphi(k)f_{k}(x)e^{-ik^{0}x},
\end{split}
\end{equation}
where $\Lambda_{\text{UV}}$ is the UV cut-off in momentum scale since we assume the  continuum description of a lattice field theory \cite{wiese2009introduction}, which naturally gives an UV cut-off, proportional to $\frac{1}{\text{lattice length scale}}$. On the other hand, $\Lambda_{0}$ is the momentum scale up to which we want description of system in terms of those interacting degrees of freedom and 
\begin{equation}
f_{k}(x)=e^{ikx}\frac{1}{m^{2}}\Big[3m^{2}\tanh^{2}(\frac{mx}{\sqrt{2}})-m^{2}-2k^{2}-3\sqrt{2}imk\tanh(\frac{mx}{\sqrt{2}})\Big].
\end{equation}
Recall that we have already taken care of $V''(\phi_{K})=\lambda(3\phi_{K}^{2}-\frac{m^{2}}{\lambda})$ in the description of normal modes. Hence, the next higher-order terms in the interacting part of the Lagrangian density are
\begin{equation}
\begin{split}
\mathcal{L}^{(3)} & =\frac{1}{3!}V'''(\phi_{K})\Phi^{3}(x)=\lambda\phi_{K}(x)\Phi^{3}(x)\\
\mathcal{L}^{(4)} & =\frac{1}{4!}V''''(\phi_{K})\Phi^{4}(x)=\frac{\lambda}{4}\Phi^{4}(x).
\end{split}
\end{equation}
First, the quadratic part of the low-energy effective action is computed in order to find the Green's function. The first contribution in $S'[\phi_{<}]$ is coming from the following 
\begin{equation}
\begin{split}
3\lambda & \int d^{2}x\phi_{<}(x)\langle\phi_{>}(x)\phi_{>}(x)\rangle_{0}\phi_{K}(x)\\
 & =3\lambda\int d^{2}x\phi_{<}(x)\phi_{K}(x)\int_{\Lambda_{0}<|k|<\Lambda_{\text{UV}}}\frac{d^{2}k_{1}}{(2\pi)^{2}}\frac{d^{2}k_{2}}{(2\pi)^{2}}\langle\varphi({k}_{1})\varphi({k}_{2})\rangle_{0}e^{i(k_{1}+k_{2})^{1}x^{1}}\tilde{f}_{k_{1}}(x)\tilde{f}_{k_{2}}(x)e^{-i(k_{1}^{0}+k_{2}^{0})x^{0}}\\
\tilde{f}_{k}(x) & =\Big[3\tanh^{2}(\frac{mx}{\sqrt{2}})-1-2\frac{k^{2}}{m^{2}}-3\sqrt{2}i\frac{k}{m}\tanh(\frac{mx}{\sqrt{2}})\Big].
\end{split}
\end{equation}
Therefore, we can write the following expression using the green function $-i\Delta_{>}(k)$ for the high-energy modes
\begin{equation}
\begin{split}
3\lambda & \int d^{2}x\phi_{<}(x)\langle\phi_{>}(x)\phi_{>}(x)\rangle_{0}\phi_{K}(x)=3\lambda\int d^{2}x\phi_{<}(x)\phi_{K}(x)\Big[\int_{\Lambda_{0}<|k|<\Lambda_{\text{UV}}}\frac{d^{2}k}{(2\pi)^{4}} -i\Delta_{>}(k)\tilde{f}_{k}(x)\tilde{f}_{-k}(x)\Big]\\
 & \approx-3\lambda\int d^{2}x\phi_{<}(x)\phi_{K}(x)\int_{\Lambda_{0}<k<\Lambda_{\text{UV}}}\frac{dk}{(2\pi)^{3}}\frac{1}{k}|\tilde{f}_{k}(x)|^{2}.
\end{split}
\end{equation}
Note that if we assume $m\ll\Lambda_{0}$ (which we did in the above last line), then we can write
\begin{equation}
\begin{split}
|\tilde{f}_{k}(x)|^{2} & =9\tanh^{4}(\frac{mx}{\sqrt{2}})-6\tanh^{2}(\frac{mx}{\sqrt{2}})+6\frac{k^{2}}{m^{2}}\tanh^{2}(\frac{mx}{\sqrt{2}})+1+4\frac{k^{4}}{m^{4}}+4\frac{k^{2}}{m^{2}},
\end{split}
\end{equation}
and as a result, we obtain
\begin{equation}
\begin{split}
\int_{\Lambda_{0}<k<\Lambda_{\text{UV}}} & \frac{dk}{(2\pi)^{3}}\frac{1}{k}|\tilde{f}_{k}(x)|^{2}\equiv g(x)=\frac{1}{(2\pi)^{3}}\Big[\ln\frac{\Lambda_{\text{UV}}}{\Lambda_{0}}(3\tanh(\frac{mx}{\sqrt{2}})-1)^{2}\\
 & +\frac{1}{m^{2}}(\Lambda_{\text{UV}}^{2}-\Lambda_{0}^{2})(2+3\tanh^{2}(\frac{mx}
{\sqrt{2}}))+\frac{1}{m^{4}}(\Lambda_{\text{UV}}^{4}-\Lambda_{0}^{4})\Big].
\end{split}
\end{equation}
Hence, as the first contribution, we obtain the following expression
\begin{equation}
S'[\phi_{<}]=-3\lambda\int d^{2}x\phi_{<}(x)\phi_{K}(x)g(x)+\ldots,
\end{equation}
and this term is proportional to $\sqrt{\lambda}$. On the other hand, the second term is equal to the following
\begin{equation}
\begin{split}
\frac{3\lambda}{2} & \int d^{2}x\langle\phi_{<}^{2}(x)\phi_{>}^{2}(x)\rangle_{0}=-\frac{3\lambda}{2}\int d^{2}x\phi_{<}^{2}(x)g(x).
\end{split}
\end{equation}
Therefore, in the leading order, we obtain
\begin{equation}
S'[\phi_{<}]=-3\lambda\int d^{2}x\phi_{<}(x)\phi_{K}(x)g(x)-\frac{3\lambda}{2}\int d^{2}x\phi_{<}^{2}(x)g(x)+\ldots.
\end{equation}
There is another contribution if we consider the second term in (\ref{6})
\begin{equation}\label{7}
\begin{split}
S'[\phi_{<}] & =-3\lambda\int d^{2}x\phi_{<}(x)\phi_{K}(x)g(x)-\frac{3\lambda}{2}\int d^{2}x\phi_{<}^{2}(x)g(x)-\frac{9i}{2}\lambda^{2}(\int d^{2}x\phi_{<}(x)\phi_{K}(x)g(x))^{2}+\ldots,
\end{split}
\end{equation}
and the third term in (\ref{6}) which leads us to our last contribution up to $\mathcal{O}(\lambda)$
\begin{equation}\label{8}
\begin{split}
\frac{9i}{2}\lambda^{2} & \int d^{2}x \ d^{2}y \ \phi_{<}(x)\phi_{<}(y)\phi_{K}(x)\phi_{K}(y)\langle\phi_{>}(x)\phi_{>}(x)\phi_{>}(y)\phi_{>}(y)\rangle_{0}\\
 & =\frac{9i}{2}\lambda^{2}\int d^{2}x \ d^{2}y \ \phi_{<}(x)\phi_{<}(y)\phi_{K}(x)\phi_{K}(y)g(x)g(y)\\
 & +9i\lambda^{2}\int d^{2}x \ d^{2}y \ \phi_{<}(x)\phi_{<}(y)\phi_{K}(x)\phi_{K}(y)
\langle\phi_{>}(x)\phi_{>}(y)\rangle_{0}^{2}.
\end{split}
\end{equation}
Note that the first term in (\ref{8}) will cancel out the last term in (\ref{7}). Further, 
\begin{equation}
\begin{split}
\langle\phi_{>}(x)\phi_{>}(y)\rangle_{0} & =\int\frac{d^{2}k}{(2\pi)^{2}}\frac{d^{2}l}{(2\pi)^{2}}
\langle\varphi(k)\varphi(l)\rangle_{0}e^{-i(k^{0}x^{0}+l^{0}y^{0})}e^{i(k^{1}x^{1}+l^{1}y^{1})}\tilde{f}_{k}(x)\tilde{f}_{l}(y)\\
 & =\int\frac{d^{2}k}{(2\pi)^{4}}-i\Delta_{>}(k)e^{-ik^{0}(x^{0}-y^{0})}e^{ik^{1}(x^{1}-y^{1})}\tilde{f}_{k}(x)\tilde{f}_{-k}(y)\\
 & =-\int_{\Lambda_{0}}^{\Lambda_{\text{UV}}}\frac{dk}{(2\pi)^{3}}e^{ik(x^{1}-x^{0}-y^{1}+y^{0})}\tilde{f}_{k}(x)\tilde{f}_{-k}(y)\frac{1}{k}\equiv\tilde{\Delta}(x,y;\Lambda_{\text{UV}},\Lambda_{0}).  
\end{split}
\end{equation}
Hence, we can write the following expression
\begin{equation}
\begin{split}
S'[\phi_{<}] & =-3\lambda\int d^{2}x\phi_{<}(x)\phi_{K}(x)g(x)-\frac{3\lambda}{2}\int d^{2}x\phi_{<}^{2}(x)g(x)\\
 & +9i\lambda^{2}\int d^{2}x \ d^{2}y \ \phi_{<}(x)\phi_{<}(y)\phi_{K}(x)\phi_{K}(y)\tilde{\Delta}^{2}(x,y;\Lambda_{\text{UV}},\Lambda_{0})\\
 & +\ldots.
\end{split}
\end{equation} 
Therefore, the quadratic part of the low-energy effective action upto $\mathcal{O}(\lambda)$ becomes the following
\begin{equation}\label{9}
\begin{split}
S_{\text{eff}}^{(\text{low})} & =S_{0}[\phi_{<}]-3\lambda\int d^{2}x\phi_{<}(x)\phi_{K}(x)g(x)-\frac{3\lambda}{2}\int d^{2}x\phi_{<}^{2}(x)g(x)\\
 & +9i\lambda^{2}\int d^{2}x \ d^{2}y \ \Big[\phi_{<}(x)\phi_{<}(y)\phi_{K}(x)\phi_{K}(y)\tilde{\Delta}^{2}(x,y;\Lambda_{\text{UV}},\Lambda_{0})\Big]+\ldots.
\end{split}
\end{equation}
Note that the imaginary contribution (similar to \cite{dunne1998qed, cheyette1985derivative}) comes from the definition of the low-energy effective action in (\ref{5.1}) according to Wilson's prescription. It is important to note that though the last term in (\ref{9}) is quadratic in field variables, it is non-local in nature. The quadratic part of the low energy effective action can also be expressed in the following form
\begin{equation}
\begin{split}
S_{\text{eff}}^{(\text{low})(2)} & =\int d^{2}x \ d^{2}y \ \phi_{<}(x)\mathbb{D}(x,y)\phi_{<}(y)\\
\mathbb{D}(x,y) & =\delta^{(2)}(x-y)\left(-\Box-\frac{1}{2}U''(\phi_{K}(x))-\frac{3\lambda}{2}g(x)\right)+9i\lambda^{2}\phi_{K}(x)\phi_{K}(y)\tilde{\Delta}^{2}(x,y;\Lambda_{\text{UV}},\Lambda_{0}).
\end{split}
\end{equation}
Note that $\tilde{\Delta}^{2}(x,y;\Lambda_{\text{UV}},\Lambda_{0})$ contributes dominantly when $(x^{0},x^{1})=(y^{0},y^{1})$ which gives
\begin{equation}\label{9.01}
\begin{split}
\mathbb{D}(x,y) & =\delta^{(2)}(x-y)\Big[-\Box-\frac{1}{2}U''(\phi_{K}(x))-\frac{3\lambda}{2}g(x)+9i\lambda^{2}\Delta\phi_{K}^{2}(x)g^{2}(x)\Big],
\end{split}
\end{equation}
where $\Delta$ is unit lattice area. However, the above differential operator is complicated enough to diagonalize or in other words, finding its normal modes and eigenvalues is difficult. Further, the eigenvalues may turn out to be imaginary since the differential operator is not hermitian in nature. Hence, we have to basically solve the following eigenvalue equation
\begin{equation}
\begin{split}
H\phi(x) & =\Big[-\frac{d^{2}}{dx^{2}}+\frac{1}{2}U''(\phi_{K}(x))+\frac{3\lambda}{2}g(x)-9i\lambda^{2}\Delta\phi_{K}^{2}(x)g^{2}(x)\Big]\phi(x)=\omega_{j}^{2}\phi(x).
\end{split} 
\end{equation}
Up to $\sech^{2}\frac{mx}{\sqrt{2}}$ term, we can write the differential operator in (\ref{9.01}) as the following
\begin{equation}\label{9.1}
\begin{split}
\mathbb{D}(x,y) & =\delta^{(2)}(x-y)\Big[-\Box-\frac{m^{2}}{2}(2-3\sech^{2}\frac{mx}{\sqrt{2}})+\frac{3\lambda a}{2}\sech^{2}\frac{mx}{\sqrt{2}}-\frac{3\lambda}{2}(a+b)+9i\lambda m^{2}\Delta(5a+b)^{2}\\
 & -9im^{2}\lambda\Delta\sech^{2}\frac{mx}{\sqrt{2}}(5a+b)^{2}\Big]\\
 & =\delta^{(2)}(x-y)\Big[-\Box-m^{2}-\frac{3\lambda}{2}(a+b)+9i\lambda m^{2}\Delta(5a+b)^{2}+\sech^{2}\frac{mx}{\sqrt{2}}\left(\frac{3}{2}(m^{2}+\lambda a)-9i\lambda\Delta m^{2}(5a+b)^{2}\right)\Big]\\
 & =\delta^{(2)}(x-y)\Big[-\Box-m^{2}-\delta_{1}^{2}+\sech^{2}\frac{mx}{\sqrt{2}}\delta_{2}^{2}\Big],
\end{split}
\end{equation}
where $a=\frac{1}{m^{2}}(\Lambda_{\text{UV}}^{2}-\Lambda_{0}^{2}), \ b=\frac{1}{m^{4}}(\Lambda_{\text{UV}}^{4}-\Lambda_{0}^{4}), \ \delta_{1}^{2}=\frac{3\lambda}{2}(a+b)-9i\lambda\Delta m^{2}(5a+b)^{2}, \ \delta_{2}^{2}=\left(\frac{3}{2}(m^{2}+\lambda a)-9i\lambda\Delta m^{2}(5a+b)^{2}\right)$. Now we discuss the reason behind not considering $\mathcal{O}(\lambda^{2})$ terms in the low-energy effective action in the next section.

\subsection{Weak coupling criterion}
During the soliton quantization, the uncertainty of momentum carried by a wavefunction increase once the scale decreases. Therefore, we want to find a small scale $L$ such that the uncertainty or the fluctuation does not diverge, and at the same time, distance scale is small \textit{w.r.t} the classical limit. This is ensured by the weak coupling criterion. The variation of a classical field around the scale $L$ is
\begin{equation}
(\Delta\phi_{L})_{\text{cl}}\sim L\frac{\partial\phi_{cl}}{\partial x}.
\end{equation}
In the $\phi^{4}$-theory case,
\begin{equation}
\begin{split}
\frac{\partial\phi_{cl}}{\partial x} & =\frac{vm}{\sqrt{2}}\sech^{2}\frac{mx}{\sqrt{2}}, \ v=\sqrt{\frac{m^{2}}{\lambda}}\\
\implies(\Delta\phi_{L})_{\text{cl}}\sim & Lmv=\frac{Lm^{2}}{\sqrt{\lambda}}.
\end{split}
\end{equation}
Let us now define a smeared out quantum field in $n+1$-dimension by
\begin{equation}
\phi_{L}(\bar{x})=\frac{1}{(2\pi L^{2})^{\frac{n}{2}}}\int d^{n}\bar{y} \ e^{-\frac{(\bar{y}-\bar{x})^{2}}{2L^{2}}}\phi(\bar{y}).
\end{equation}
The fluctuation of the above quantum field around the vacuum is given by
\begin{equation}
\begin{split}
(\Delta\phi_{L})_{\text{quantum}}^{2} & \equiv\bra{0}\phi_{L}^{2}(0)\ket{0}=\frac{1}{(2\pi L^{2})^{n}}\int d^{n}\bar{y}d^{n}\bar{z} \ e^{-\frac{(\bar{y}^{2}+\bar{z}^{2})}{2L^{2}}}\bra{0}\phi(\bar{y})\phi(\bar{z})\ket{0}.
\end{split}
\end{equation}
Using the mode expansion formula for free field operators, $\bra{0}\phi(\bar{y})\phi(\bar{z})\ket{0}$ can easily be evaluated which is the following
\begin{equation}
\bra{0}\phi(\bar{y})\phi(\bar{z})\ket{0}=\int\frac{d^{n}p}{(2\pi)^{n}}\frac{1}{2\sqrt{p^{2}+m^{2}}}e^{ip.(\bar{y}-\bar{z})}.
\end{equation}
This leads to the following expression of fluctuation
\begin{equation}\label{10}
\begin{split}
(\Delta\phi_{L})_{\text{quantum}}^{2} & =\frac{1}{(2\pi L)^{2n}}\int d^{n}\bar{y}d^{n}\bar{z}d^{n}p\frac{1}{2\sqrt{p^{2}+m^{2}}}e^{-\frac{(\bar{y}^{2}+\bar{z}^{2})}{2L^{2}}}e^{ip.(\bar{y}-\bar{z})}=\int\frac{d^{n}p}{(2\pi)^{n}}\frac{e^{-p^{2}L^{2}}}{2\sqrt{p^{2}+m^{2}}}.
\end{split}
\end{equation}
We know that the variation of the kink is more around center about $m^{-1}$ scale, hence, the length scale $L$ must satisfy
\begin{equation}
L\ll m^{-1}\implies mL\ll 1.
\end{equation}
As a result, the integration in (\ref{10}) gives the following result
\begin{equation}
(\Delta\phi_{L})_{\text{quantum}}\sim\sqrt{\ln(\frac{1}{mL})}, \ \text{if} \ n=1.
\end{equation} 
For solitons in $1+1$ dimension case, we can now set up the weak coupling criterion. We want the classical fluctuations to be a lot larger than the quantum fluctuations which means the following inequality
\begin{equation}
\sqrt{\ln(\frac{1}{mL})}\ll\frac{m^{2}L}{\sqrt{\lambda}}\implies 1 \ll mL \ e^{\frac{m^{4}L^{2}}{\lambda}},
\end{equation}
must hold. Hence, the weak coupling restricts $\lambda$ to be very small even if $mL\ll 1$. Because of this reason, we only keep the terms in the low energy effective action whose strength is $\mathcal{O}(\lambda)$ in nature.

\subsection{Low-energy effective action in large length scale}
In this section, we show the dissipative nature of the low-energy excitations, which is our third result. The differential operator in last line of (\ref{9.1}) in long range $x\gg \frac{m^{-1}}{2}$ or $x\ll-\frac{m^{-1}}{2}$ effectively becomes the following
\begin{equation}
\mathbb{D}(x,y)=\delta^{(2)}(x-y)\Big[-\Box-m^{2}-\delta_{1}^{2}\Big],
\end{equation} 
and the action effectively becomes
\begin{equation}\label{11}
\begin{split}
S_{\text{eff}}^{(\text{low})}[\phi_{<}] & =\int d^{2}x \ \Bigg[\phi_{<}(x)\Big[-\Box-m^{2}-\delta_{1}^{2}\Big]\phi_{<}(x)-3\lambda\delta_{3}\epsilon(x^{1})\phi_{<}(x)+m\sqrt{\lambda}\epsilon(x^{1})\phi_{<}^{3}(x)+\frac{\lambda}{4}\phi_{<}^{4}(x)
\Bigg],
\end{split}
\end{equation}
where $\int d^{2}x\equiv\int_{-\infty}^{\infty}dt\left(\int_{x_{1}}^{\infty}dx+\int_{-\infty}^{-x_{1}}dx\right)$, $x_{1}\gg m^{-1}, \ \delta_{3}=3a+b$ and $\epsilon(x)=\theta(x)-\theta(-x)$. Further, we impose following condition $\Lambda_{0}x_{1}\ll1$ which leads to the following approximation
\begin{equation}
\begin{split}
\Big[\int_{x_{1}}^{\infty}dx & +\int_{-\infty}^{-x_{1}}dx\Big]e^{i(k^{1}+l^{1}).x}=\Big[e^{-i(k^{1}+l^{1})x_{1}}\int_{0}^{\infty}dx+e^{i(k^{1}+l^{1}).x_{1}}\int_{-\infty}^{0}dx\Big]e^{i(k^{1}+l^{1}).x}\\
\approx\int_{-\infty}^{\infty}dx & \ e^{i(k^{1}+l^{1}).x}=(2\pi)\delta(k^{1}+l^{1})
\end{split}.
\end{equation}
As a result of the above approximation, we can effectively write the quadratic term in momentum space as follows
\begin{equation}
S_{\text{eff}}^{(\text{low})(2)}[\phi_{<}]=\int_{-\infty}^{\infty}\frac{dk^{0}}{2\pi}\int_{-\Lambda_{0}}^{\Lambda_{0}}\frac{dk^{1}}{2\pi}\phi_{<}(-k)\Big[k^{2}-m^{2}-\delta_{1}^{2}\Big]\phi_{<}(k).
\end{equation}
Hence, in the long range, the Green function in the free-field theory becomes
\begin{equation}
\mathcal{G}(x,y)=\int_{-\infty}^{\infty}\frac{dk^{0}}{2\pi}\int_{-\Lambda_{0}}^{\Lambda_{0}}\frac{dk^{1}}{2\pi}\frac{1}{k^{2}-m^{2}-\delta_{1}^{2}}e^{-ik.(x-y)},
\end{equation}
where $k^{2}=(k^{0})^{2}-(k^{1})^{2}$. Note that the poles of the Green's function in free-field theory are at $k^{0}=\pm\sqrt{(k^{1})^{2}+m^{2}+\delta_{1}^{2}}$ where $\delta_{1}^{2}$ is a complex number with negative imaginary term. This dispersion relation shows the dissipative effect in the system under the time evolution. This shows that the short-range correlations (because of the presence of mass term) of the excitations do not last too long because of the dissipative effect.

In the similar way, the fourth term in (\ref{11}) can also be written in the following way
\begin{equation}
\begin{split}
\int_{x}\phi_{<}^{4}(x) & =\int_{k_{1}}\int_{k_{2}}\int_{k_{3}}\int_{k_{4}}\delta^{(2)}(k_{1}+k_{2}+k_{3}+k_{4})\phi_{<}(k_{1})\phi_{<}(k_{2})\phi_{<}(k_{3})\phi_{<}(k_{4}).
\end{split}
\end{equation}
On the other hand, using the following result
\begin{equation}
\begin{split}
-e^{ika}\int_{-\infty}^{0}dx \ e^{ik.x} & +e^{-ika}\int_{0}^{\infty}dx \ e^{ik.x}\\
=-e^{ika}\int_{-\infty(1-i\epsilon)}^{0}dx \ e^{ik.x} & +e^{-ika}\int_{0}^{\infty(1+i\epsilon)}dx \ e^{ik.x}=\frac{2i\cos(ka)}{k}\approx\frac{2i}{k}, \ \text{if} \ ka\ll1,
\end{split}
\end{equation} 
the low-energy effective action in momentum space can be expressed as follows
\begin{equation}
\begin{split}
S_{\text{eff}}^{(\text{low})}[\phi_{<}] & =\int_{k}\phi_{<}(-k)\Big[k^{2}-m^{2}-\delta_{1}^{2}\Big]\phi_{<}(k)-6i\lambda\delta_{3}\int_{k}\frac{\phi_{<}(k)}{k^{1}}\\
 & +2im\sqrt{\lambda}\int_{k_{1}}\int_{k_{2}}\int_{k_{3}}\Big[\frac{1}{k_{1}^{1}+k_{2}^{1}+k_{3}^{1}}\phi_{<}(k_{1})\phi_{<}(k_{2})\phi_{<}(k_{3})\Big]\\
 & +\lambda\int_{k_{1}}\int_{k_{2}}\int_{k_{3}}\int_{k_{4}}\delta^{(2)}(k_{1}+k_{2}
+k_{3}+k_{4})\phi_{<}(k_{1})\phi_{<}(k_{2})\phi_{<}(k_{3})\phi_{<}(k_{4}),
\end{split}
\end{equation}
where $\int_{k}=\int_{-\infty}^{\infty}\frac{dk^{0}}{2\pi}\int_{-\Lambda_{0}}^{\Lambda_{0}}\frac{dk^{1}}{2\pi}$.

It is important here to emphasize that the presence of the second term in the above action behaves as a source term that generates a one-point function even in the free-field theory. This shows that the expectation value of fluctuations is non-zero even in the free-field theory. This feature can be used to look for the presence of soliton-like classical configurations in the system unlike, many-body field theory about the trivial vacuum in which 1-point function vanishes identically. This is also a feature, found at the level of quantum corrections by integrating out the high-energy degrees of freedom from the system.

In order to avoid the IR(infra-red) divergences, from now on we put an IR cut-off in the momentum domain integration denoted by $\Lambda_{\text{IR}}$. This arises anyway once we choose a finite-size system. Hence, we fix the domain of integration in $\Lambda_{\text{IR}}<|k^{1}|<\Lambda_{0}$. 

\section{Importance of collective coordinates in low-energy effective field theory}
Scattering of solitons in field theory from defects and impurities have been studied mainly using numerical analysis. However, some analytical results are shown using a mathematical technique, known as the collective coordinates \cite{takyi2016collective, weigel2019collective, burzlaff1998soliton, nazifkar2010collective}. Within some approximations, the predictions coming from the collective coordinates are very precise. In this section, we present a brief review of collective coordinates through a simple example, then we discuss the usefulness of this method in the present context.

Let us consider a Sine-Gordon theory with the following action
\begin{equation}\label{CC1}
\mathcal{A}=\int d^{2}x\Big[\frac{1}{2}\partial_{\mu}\phi\partial^{\mu}\phi-\lambda(x)(1-\cos\phi)\Big], \ \lambda(x)=\lambda_{0}+V(x),
\end{equation}  
where $V(x)$ is an external potential. For an arbitrary potential $V(x)$, the Euler-Lagrange equation
\begin{equation}\label{EL2}
\partial_{\mu}\partial^{\mu}\phi+\lambda(x)\sin\phi=0,
\end{equation}
cannot be solved analytically. Since $V(x)=0$ leads us to the original Sine-Gordon theory with the well-known solutions of kink and anti-kink solitons, solution of the equation (\ref{EL2}) is expected to be of the following form
\begin{equation}
\phi(x,X(t))=4\taninv\left(e^{\sqrt{\lambda_{0}}\frac{x-X(t)}{\sqrt{1-\dot{X}^{2}}}}\right).
\end{equation}
Inserting the above ansatz into the action (\ref{CC1}) with the adiabatic approximation \cite{kivshar1991resonant, fei1992resonant}, we obtain the following Lagrangian density
\begin{equation}\label{CC2}
\mathcal{L}=2(\dot{X}^{2}-1)\sech^{2}(x-X(t))-2\lambda(x)\sech^{2}(x-X(t)).
\end{equation}
Integrating the above Lagrangian density \textit{w.r.t} the spatial coordinate and considering $\lambda_{0}=1$, we obtain the following Lagrangian
\begin{equation}\label{L1}
L[X,\dot{X}]=4\dot{X}^{2}-8-2\int V(x)\sech^{2}(x-X(t))dx.
\end{equation}
One of the simple and interesting physical examples of such potentials is Dirac-delta potential, often used for scatterings. In this case, $V(x)=\epsilon\delta(x)$ which makes the Lagrangian (\ref{L1}) as
\begin{equation}
L[X,\dot{X}]=4\dot{X}^{2}-8-2\epsilon\sech^{2}(X).
\end{equation} 
The equation of motion in the variable $X(t)$ is 
\begin{equation}
8\ddot{X}-4\epsilon\sech^{2}(X)\tanh(X)=0,
\end{equation}
and the general solution of the above equation is of the following form
\begin{equation}
\dot{X}^{2}=\frac{1}{2}\epsilon\sech^{2}(X_{0})+\dot{X}_{0}^{2}-\frac{1}{2}\epsilon\sech^{2}(X),
\end{equation}
where $X_{0}$ and $\dot{X}_{0}$ are the initial conditions at $t=0$. It is also important to emphasize that $\epsilon>0$ creates a barrier and $\epsilon<0$ creates a potential well.

The above mathematical method can be summarized in the following way that can be generalized to the coupled soliton models. The internal structure or more specifically spatial structure of solitons can be omitted by integrating the Lagrangian density over the spatial coordinates. This integrated Lagrangian density is known as the collective Lagrangian. Although after the integration, solitons appear as point-like particles, however, the effect of the spatial extension of solitons reflects from the kinetic terms and the potential terms of the collective Lagrangian, as shown in the above example.

The low-energy effective action, shown in (\ref{9}) or its momentum space version (\ref{11}) shows explicitly the dependence of different external potential in the action due to integrating out the high-energy modes. Hence, the collective coordinates would be useful to find out the dynamics of low-energy excitations from the low-energy effective action. Further, the factor $\epsilon(x^{1})$ in (\ref{11}) shows the potential barrier in the spatial coordinate $x^{1}$ which plays important role in scattering phenomena. Thus, the collective coordinates open out a new domain for studying low-energy excitations about the soliton configurations.
  
\section{Physical Observables}\label{section4}
In QFT, every observable for example S-matrix \cite{bugrij1995s, conde2014physics} can be achieved from the set of all the correlation or $n$-point functions \cite{cardy2010introduction} for various values of positive integer $n$. Therefore, in the following section, we compute a few correlation functions.

\subsection{One-point function}
Let us first assume the low-energy effective action up to the quadratic terms in field variables. Hence, we can write the generating functional as follows
\begin{equation}
\begin{split}
\mathcal{Z}_{\text{free}}[j(x)] & =\int\mathcal{D}\phi_{<}e^{iS^{(2)}[\phi_{<}]+i\int d^{2}x j(x)\phi_{<}(x)}, \ S^{(2)}[\phi_{<}]=\int_{k}\phi_{<}(-k)\Big[k^{2}-m^{2}-\delta_{1}^{2}\Big]\phi_{<}(k)-6i\lambda\delta_{3}\int_{k}\frac{\phi_{<}(k)}{k^{1}}\\
\implies\mathcal{Z}_{\text{free}}[j(x)] & =\mathcal{Z}_{\text{free}}[0]e^{i\int_{k}(j(-k)+\frac{6i\lambda}{k^{1}})\mathcal{G}_{F}(k)(j(k)-\frac{6i\lambda}{k^{1}})}e^{-i\int_{k}\frac{36\lambda^{2}\mathcal{G}_{F}(k)}{(k^{1})^{2}}}\\
\implies\mathcal{Z}_{\text{free}}[j(x)] & =\mathcal{Z}_{\text{free}}[0] e^{\int_{k}\Big[i j(-k)\mathcal{G}_{F}(k)j(k)+j(-k)\mathcal{G}_{F}(k)\frac{6\lambda}{k^{1}}-j(k)\mathcal{G}_{F}(k)\frac{6\lambda}{k^{1}}\Big]}, \ \mathcal{G}_{F}(k)=\frac{1}{k^{2}-m^{2}-\delta_{1}^{2}}.
\end{split}
\end{equation} 
By definition, the connected one-point function is given by the following
\begin{equation}
\begin{split}
\langle\phi_{<}(k)\rangle_{c} & =\frac{\delta}{i\delta j(-k)}\ln\mathcal{Z}_{\text{free}}[j]\Big|_{j=0}=\frac{1}{\mathcal{Z}_{\text{free}}[0]}\frac{\delta\mathcal{Z}_{\text{free}}[j]}{i\delta j(-k)}\Big|_{j=0}=-\frac{12i\lambda}{k^{1}}\mathcal{G}_{F}(k)=-\frac{12i\lambda}{k^{1}(k^{2}-m^{2}-\delta_{1}^{2})}\\
\implies\langle\phi_{<}(x)\rangle_{c} & =\int_{k}e^{-ik.x}\langle\phi_{<}(k)\rangle_{c}=-12i\lambda\int_{k}e^{-ik.x}\frac{1}{k^{1}(k^{2}-m^{2}-\delta_{1}^{2})}.
\end{split}
\end{equation}
Since $\delta_{1}^{2}$ lies in the lower half of the complex plane, therefore, $\sqrt{(k^{1})^{2}+m^{2}+\delta_{1}^{2}}$ also does lie in the lower half of the complex plane. Using this fact, we can write
\begin{equation}
\begin{split}
\langle\phi_{<}(x)\rangle_{c} & =-6\lambda\int_{\Lambda_{\text{IR}}}^{\Lambda_{0}}\frac{dk^{1}}{2\pi}\frac{1}{k^{1}\sqrt{(k^{1})^{2}+m^{2}+\delta_{1}^{2}}}\Big[e^{ik^{1}x-i\sqrt{(k^{1})^{2}+m^{2}+\delta_{1}^{2}}t}\theta(t)-e^{i(k^{1}x+\sqrt{(k^{1})^{2}+m^{2}+\delta_{1}^{2}}t)}\theta(-t)\Big]\\
 & =-6\lambda\varepsilon(t)\int_{\Lambda_{\text{IR}}}^{\Lambda_{0}}\frac{dk^{1}}{2\pi}\Big[\frac{1}{k^{1}\sqrt{(k^{1})^{2}+m^{2}+\delta_{1}^{2}}}e^{i(k^{1}x-\sqrt{(k^{1})^{2}+m^{2}+\delta_{1}^{2}}|t|)}\Big].
\end{split}
\end{equation}
Since $\sqrt{(k^{1})^{2}+m^{2}+\delta_{1}^{2}}$ lies in the lower half of the complex plane, $<\phi_{<}(x)>_{c}$ decays exponentially in time as $t\rightarrow\pm\infty$. Hence, in the large length scale in free-field theory, the one-point function vanishes in the asymptotic time limit. On the other hand, in the full interacting theory, the generating function takes the following form
\begin{equation}
\begin{split}
\mathcal{Z}[j] & =\mathcal{Z}_{\text{free}}[0]e^{i\lambda\int_{k_{1}}\int_{k_{2}}\int_{k_{3}}\int_{k_{4}}\delta^{(2)}(\sum_{i=1}^{4}k_{i})\frac{\delta}{\delta j(-k_{1})}\frac{\delta}{\delta j(-k_{2})}\frac{\delta}{\delta j(-k_{3})}\frac{\delta}{\delta j(-k_{4})}}\\
 & \times e^{-2m\sqrt{\lambda}\int_{k_{1}}\int_{k_{2}}\int_{k_{3}}\frac{1}{k_{1}^{1}+k_{2}^{1}+k_{3}^{1}}\frac{\delta}{i\delta j(-k_{1})}\frac{\delta}{i\delta j(-k_{2})}\frac{\delta}{i\delta j(-k_{3})}}e^{\int_{k}\Big[i j(-k)\mathcal{G}_{F}(k)j(k)+j(-k)\mathcal{G}_{F}(k)\frac{6\lambda}{k^{1}}-j(k)\mathcal{G}_{F}(k)\frac{6\lambda}{k^{1}}\Big]}.
\end{split}
\end{equation}
Then using the following definition
\begin{equation}\label{12}
\begin{split}
\langle & \phi_{<}(k)\rangle=\frac{\delta}{i\delta j(-k)}e^{i\lambda\int_{k_{1}}\int_{k_{2}}\int_{k_{3}}\int_{k_{4}}\delta^{(2)}(\sum_{i=1}^{4}k_{i})\frac{\delta}{\delta j(-k_{1})}\frac{\delta}{\delta j(-k_{2})}\frac{\delta}{\delta j(-k_{3})}\frac{\delta}{\delta j(-k_{4})}}\\
 & \times e^{-2m\sqrt{\lambda}\int_{k_{1}}\int_{k_{2}}\int_{k_{3}}\frac{1}{k_{1}^{1}+k_{2}^{1}+k_{3}^{1}}\frac{\delta}{i\delta j(-k_{1})}\frac{\delta}{i\delta j(-k_{2})}\frac{\delta}{i\delta j(-k_{3})}}e^{\int_{k}\Big[i j(-k)\mathcal{G}_{F}(k)j(k)+j(-k)\mathcal{G}_{F}(k)\frac{6\lambda}{k^{1}}-j(k)\mathcal{G}_{F}(k)\frac{6\lambda}{k^{1}}\Big]}\Bigg|_{j=0},
\end{split}
\end{equation}
we obtain the following expression
\begin{equation}
\begin{split}
\langle\phi_{<}(k)\rangle  & =\langle\phi_{<}(k)\rangle_{c}^{(0)}-2m\sqrt{\lambda}\int_{k_{1}}\int_{k_{2}}\int_{k_{3}}\frac{1}{k_{1}^{1}+k_{2}^{1}+k_{3}^{1}}\Big[\frac{\delta}{i\delta j(-k_{1})}\frac{\delta}{i\delta j(-k_{2})}\frac{\delta}{i\delta j(-k_{3})}\frac{\delta}{i\delta j(-k)}\Big]\\
 & \times e^{\int_{k}\Big[i j(-k)\mathcal{G}_{F}(k)j(k)+j(-k)\mathcal{G}_{F}(k)\frac{6\lambda}{k^{1}}-j(k)\mathcal{G}_{F}(k)\frac{6\lambda}{k^{1}}\Big]}\Bigg|_{j=0}+\ldots.
\end{split}
\end{equation}
After doing the functional differentiation, we obtain
\begin{equation}\label{12.1}
\begin{split}
<\phi_{<}(k)> & =<\phi_{<}(k)>_{c}^{(0)}-2m\sqrt{\lambda}\Bigg[-\frac{12}{k^{1}}\mathcal{G}_{F}(k)\int_{k_{1}}\mathcal{G}_{F}(k_{1})\Big[1+24\lambda^{2}\int_{k_{3}}\frac{\mathcal{G}_{F}(k_{3})}{(k_{3}^{1})^{2}}\Big]\\
-288\lambda & \Big[\int_{k_{2}}\int_{k_{3}}\frac{\mathcal{G}_{F}(k_{2})}{k_{2}^{1}}\frac{\mathcal{G}_{F}(k_{3})}{k_{3}^{1}}\frac{\mathcal{G}_{F}(k)}{k_{2}^{1}+k_{3}^{1}-k^{1}}+\frac{\mathcal{G}_{F}(k)}{k^{1}}\int_{k_{2}}\int_{k_{1}}\mathcal{G}_{F}(k_{1})\frac{\mathcal{G}_{F}(k_{2})}{(k_{2}^{1})^{2}}\Big]\\
 & -288\lambda^{2}\int_{k_{1}}\int_{k_{3}}\frac{\mathcal{G}_{F}(k_{1})}{k_{1}^{1}}\frac{\mathcal{G}_{F}(k_{3})}{k_{3}^{1}}\frac{\mathcal{G}_{F}(k)}{k_{1}^{1}+k_{3}^{1}-k^{1}}-288\lambda^{2}\frac{\mathcal{G}_{F}(k)}{k^{1}}\int_{k_{1}}\int_{k_{3}}\frac{\mathcal{G}_{F}(k_{1})}{(k_{1}^{1})^{2}}\mathcal{G}_{F}(k_{3})\\
-144\lambda^{2} & \int_{k_{1}}\int_{k_{2}}\frac{\mathcal{G}_{F}(k_{1})}{k_{1}^{1}}\frac{\mathcal{G}_{F}(k_{2})}{k_{2}^{1}}\frac{\mathcal{G}_{F}(k)}{k_{1}^{1}+k_{2}^{1}-k^{1}}+(144)^{2}\lambda^{4}\frac{\mathcal{G}_{F}(k)}{k^{1}}\int_{k_{1}}\int_{k_{2}}\int_{k_{3}}\frac{1}{k_{1}^{1}k_{2}^{1}k_{3}^{1}(k_{1}^{1}+k_{2}^{1}+k_{3}^{1})}\\
 & \times\mathcal{G}_{F}(k_{1})\mathcal{G}_{F}(k_{2})\mathcal{G}_{F}(k_{3})\Bigg]+\ldots\\
 & =<\phi_{<}(k)>^{(0)}+\frac{24m\sqrt{\lambda}}{k^{1}}\mathcal{G}_{F}(k)\int_{k_{1}}\mathcal{G}_{F}(k_{1})+576m\sqrt{\lambda}\lambda\Big[\int_{k_{2}}\int_{k_{3}}\frac{\mathcal{G}_{F}(k_{2})}{k_{2}^{1}}\frac{\mathcal{G}_{F}(k_{3})}{k_{3}^{1}}\frac{\mathcal{G}_{F}(k)}{k_{2}^{1}+k_{3}^{1}-k^{1}}\\
 & +\frac{\mathcal{G}_{F}(k)}{k^{1}}\int_{k_{2}}\int_{k_{1}}\mathcal{G}_{F}(k_{1})\frac{\mathcal{G}_{F}(k_{2})}{(k_{2}^{1})^{2}}\Big]+\mathcal{O}(\lambda^{2}), 
\end{split}
\end{equation}
where
\begin{equation}
\begin{split}
\int_{k_{1}}\mathcal{G}_{F}(k_{1}) & =-\frac{i}{2\pi}\Big[\sinh^{-1}\frac{\Lambda_{0}^{2}}{m^{2}+\delta_{1}^{2}}-\sinh^{-1}\frac{\Lambda_{\text{IR}}^{2}}{m^{2}+\delta_{1}^{2}}\Big]\\
\int_{k_{2}}\frac{\mathcal{G}_{F}(k_{2})}{(k_{2}^{1})^{2}} & =-\frac{i}{2\pi(m^{2}+\delta_{2}^{2})}\Big[\coth\left(\sinh^{-1}\frac{\Lambda_{\text{IR}}^{2}}{m^{2}+\delta_{1}^{2}}\right)-\coth\left(\sinh^{-1}\frac{\Lambda_{0}^{2}}{m^{2}+\delta_{1}^{2}}\right)\Big]\\
\int_{k_{1}}\int_{k_{2}}\frac{\mathcal{G}_{F}(k_{1})}{k_{1}^{1}} & \frac{\mathcal{G}_{F}(k_{2})}{k_{2}^{1}}\frac{1}{k_{1}^{1}+k_{2}^{1}-k^{1}}=-\int_{k_{1}}\int_{k_{2}}\frac{\mathcal{G}_{F}(k_{1})}{k_{1}^{1}}\frac{\mathcal{G}_{F}(k_{2})}{k_{2}^{1}}\frac{1}{k_{1}^{1}+k_{2}^{1}+k^{1}}.
\end{split}
\end{equation}
Therefore, if we define $f(k^{1})\equiv\int_{k_{1}}\int_{k_{2}}\frac{\mathcal{G}_{F}(k_{1})}{k_{1}^{1}} \frac{\mathcal{G}_{F}(k_{2})}{k_{2}^{1}}\frac{1}{k_{1}^{1}+k_{2}^{1}-k^{1}}$, then $f(k^{1})=-f(-k^{1})$, which also means $f(0)=0$. Further, we can write the following Taylor-series expansion
\begin{equation}
\begin{split}
f(k^{1}) & =k^{1}f^{(1)}(0)+\frac{1}{3!}(k^{1})^{3}f^{(3)}(0)+\ldots=k^{1}\left(f^{(1)}(0)+\frac{1}{3!}(k^{1})^{2}f^{(3)}(0)+\ldots\right),
\end{split}
\end{equation}
where
\begin{equation}
\begin{split}
f^{(1)}(0) & =\int_{k_{1}}\int_{k_{2}}\frac{\mathcal{G}_{F}(k_{1})}
{k_{1}^{1}}\frac{\mathcal{G}_{F}(k_{2})}{k_{2}^{1}}\frac{1}{(k_{1}^{1}+k_{2}^{1})^{2}}\\
 & =-\left(\int_{\Lambda_{\text{IR}}}^{\Lambda_{0}}+\int_{-\Lambda_{0}}^{-\Lambda_{\text{IR}}}\right)\frac{dk}{2\pi}\left(\int_{\Lambda_{\text{IR}}}^{\Lambda_{0}}+\int_{-\Lambda_{0}}^{-\Lambda_{\text {IR}}}\right)\frac{dl}{2\pi}\frac{1}{4\sqrt{k^{2}+m^{2}+\delta_{1}^{2}}\sqrt{l^{2}+m^{2}+\delta_{1}^{2}}}\frac{1}{kl(k+l)^{2}}\\
 & =2\int_{\Lambda_{\text{IR}}}^{\Lambda_{0}}\int_{\Lambda_{\text{IR}}}^{\Lambda_{0}} \frac{dk}{2\pi}\frac{dl}{2\pi}\Big[\frac{1}{\sqrt{k^{2}+m^{2}+\delta_{1}^{2}}\sqrt{l^{2}+m^{2}+\delta_{1}^{2}}}\frac{1}{(k^{2}-l^{2})^{2}}\Big]. 
\end{split}
\end{equation}
The above integration can be done numerically. Once we know the above term for small $k^{1}\in(\Lambda_{\text{IR}},\Lambda_{0})$, we can write $f(k^{1})=f^{(1)}(0)k^{1}$ since we are considering the low-energy regime.

There is another contribution of the $\mathcal{O}(\lambda)$ coming from four functional derivatives of action which is there in the exponent in (\ref{12}). However, the action of functional derivative operators only on the $\lambda$ independent term in the last exponent of (\ref{12}) gives the following
\begin{equation}
\begin{split}
i\lambda & \frac{\delta}{i\delta j(-k)}\int_{k_{1}}\int_{k_{2}}\int_{k_{3}}\int_{k_{4}}\delta^{(2)}(k_{1}+k_{2}+k_{3}+k_{4})\frac{\delta}{i\delta j(-k_{1})}\frac{\delta}{i\delta j(-k_{2})}\frac{\delta}{i\delta j(-k_{3})}\frac{\delta}{i\delta j(-k_{4})}e^{i\int_{k}j(-k)\mathcal{G}_{F}(k)j(k)}\Bigg|_{j=0}.
\end{split}
\end{equation}
The above contribution is trivially zero. A similar thing happens for another $\lambda\sqrt{\lambda}$ order contribution which comes from the seven functional derivatives acting on the same exponent and it gives rise to zero contribution. Hence, our result for one-point function up to order $\mathcal{O}(\lambda\sqrt{\lambda})$ is correct.
At last, we want to emphasize that because of the presence of pole of $k^{0}$ in the complex plane, the one-point function of these fluctuations is exponentially decaying even in interacting theory, which can be solved perturbatively.

\subsection{Two-point function}
The expression of two-point function up to the normalization factor $\mathcal{Z}_{\text{free}}[0]$ is the following
\begin{equation}
\begin{split}
\langle\phi_{<}(k_{1})\phi_{<}(k_{2})\rangle & =\frac{1}{\mathcal{Z}_{\text{free}}[0]}\frac{\delta}{i\delta j(-k_{1})}\frac{\delta}{i\delta j(-k_{2})}\mathcal{Z}[j]\Big|_{j=0}=<\phi_{<}(k_{1})\phi_{<}(k_{2})>_{0}+\ldots,
\end{split}
\end{equation}
where $\ldots$ denotes the higher functional derivative contribution, which is discussed  later. Let us first calculate $<\phi_{<}(k_{1})\phi_{<}(k_{2})>_{0}$, which is given by
\begin{equation}
\begin{split}
\langle\phi_{<}(k_{1}) & \phi_{<}(k_{2})\rangle_{0}=\frac{\delta}{i\delta j(-k_{1})}\Big[j(k_{2})\mathcal{G}_{F}(k_{2})-12i\mathcal{G}_{F}(k_{2})\frac{\lambda}{k_{2}^{1}}\Big]e^{\int_{k}\Big[i j(-k)\mathcal{G}_{F}(k)j(k)+j(-k)\mathcal{G}_{F}(k)\frac{6\lambda}{k^{1}}-j(k)\mathcal{G}_{F}(k)\frac{6\lambda}{k^{1}}\Big]}\Bigg|_{j=0}\\
 & =-i\delta^{(2)}(k_{1}+k_{2})\mathcal{G}_{F}(k_{1})-\frac{\mathcal{G}_{F}(k_{1})\mathcal{G}_{F}(k_{2})}{k_{1}^{1}k_{2}^{1}}144\lambda^{2}.
\end{split}
\end{equation}
Now onwards, we keep the terms only up to $\mathcal{O}(\lambda^{2})$.

Now we write down $\mathcal{O}(\lambda)$ contribution which is the following
\begin{equation}
\begin{split}
-8\lambda & \int_{l_{1}}\int_{l_{2}}\int_{l_{3}}\int_{l_{4}}\delta^{(2)}(l_{1}+l_{2}+l_{3}+l_{4})\Big[\delta^{(2)}(l_{4}+l_{2})\delta^{(2)}(l_{3}+l_{1})\delta^{(2)}(k_{1}+k_{2})\mathcal{G}_{F}(l_{4})\mathcal{G}_{F}(l_{3})\mathcal{G}_{F}(k_{1})\\
+\delta^{(2)} & (l_{2}+l_{3})\delta^{(2)}(l_{4}+l_{1})\delta^{(2)}(k_{1}+k_{2})\mathcal{G}_{F}(l_{4})\mathcal{G}_{F}(l_{3})\mathcal{G}_{F}(k_{1})+\delta^{(2)}(l_{2}+l_{1})\delta^{(2)}(l_{4}+l_{3})\delta^{(2)}(k_{1}+k_{2})\mathcal{G}_{F}(l_{4})\mathcal{G}_{F}(l_{3})\mathcal{G}_{F}(k_{1})\\
+\delta^{(2)} & (l_{4}+l_{1})\delta^{(2)}(l_{3}+k_{1})\delta^{(2)}(l_{2}+k_{2})\mathcal{G}_{F}(l_{4})\mathcal{G}_{F}(l_{3})\mathcal{G}_{F}(l_{2})+\delta^{(2)}(l_{4}+l_{1})\delta^{(2)}(l_{3}+k_{2})\delta^{(2)}(l_{2}+k_{1})\mathcal{G}_{F}(l_{4})\mathcal{G}_{F}(l_{3})\mathcal{G}_{F}(l_{2})\\
+\delta^{(2)} & (l_{3}+l_{1})\delta^{(2)}(l_{4}+k_{2})\delta^{(2)}(l_{2}+k_{1})\mathcal{G}_{F}(l_{4})\mathcal{G}_{F}(l_{3})\mathcal{G}_{F}(l_{2})+\delta^{(2)}(l_{3}+l_{1})\delta^{(2)}(l_{4}+k_{1})\delta^{(2)}(l_{2}+k_{2})\mathcal{G}_{F}(l_{4})\mathcal{G}_{F}(l_{3})\mathcal{G}_{F}(l_{2})\\
+\delta^{(2)} & (l_{2}+l_{1})\delta^{(2)}(l_{4}+k_{2})\delta^{(2)}(l_{3}+k_{1})\mathcal{G}_{F}(l_{4})\mathcal{G}_{F}(l_{3})\mathcal{G}_{F}(l_{2})+\delta^{(2)}(l_{2}+l_{1})\delta^{(2)}(l_{4}+k_{1})\delta^{(2)}(l_{3}+k_{2})\mathcal{G}_{F}(l_{4})\mathcal{G}_{F}(l_{3})\mathcal{G}_{F}(l_{2})\\
+\delta^{(2)} & (l_{2}+l_{4})\delta^{(2)}(l_{1}+k_{1})\delta^{(2)}(l_{3}+k_{2})\mathcal{G}_{F}(l_{4})\mathcal{G}_{F}(l_{3})\mathcal{G}_{F}(l_{1})+\delta^{(2)}(l_{2}+l_{4})\delta^{(2)}(l_{1}+k_{2})\delta^{(2)}(l_{3}+k_{1})\mathcal{G}_{F}(l_{4})\mathcal{G}_{F}(l_{3})\mathcal{G}_{F}(l_{1})\\
+\delta^{(2)} & (l_{2}+l_{3})\delta^{(2)}(l_{4}+k_{1})\delta^{(2)}(l_{1}+k_{2})\mathcal{G}_{F}(l_{4})\mathcal{G}_{F}(l_{3})\mathcal{G}_{F}(l_{1})+\delta^{(2)}(l_{2}+l_{3})\delta^{(2)}(l_{4}+k_{2})\delta^{(2)}(l_{1}+k_{1})\mathcal{G}_{F}(l_{4})\mathcal{G}_{F}(l_{3})\mathcal{G}_{F}(l_{1})\\
+\delta^{(2)} & (l_{4}+l_{3})\delta^{(2)}(l_{2}+k_{2})\delta^{(2)}(l_{1}+k_{1})\mathcal{G}_{F}(l_{4})\mathcal{G}_{F}(l_{2})\mathcal{G}_{F}(l_{1})+\delta^{(2)}(l_{4}+l_{3})\delta^{(2)}(l_{2}+k_{1})\delta^{(2)}(l_{1}+k_{2})\mathcal{G}_{F}(l_{4})\mathcal{G}_{F}(l_{2})\mathcal{G}_{F}(l_{1})\Big].
\end{split}
\end{equation}
This can be further simplified as the following
\begin{equation}
\begin{split}
-8\lambda & \Big[2\delta^{(2)}(0)\delta^{(2)}(k_{1}+k_{2})\mathcal{G}_{F}(k_{1})\left(\int_{l}\mathcal{G}_{F}(l)\right)^{2}+\delta^{(2)}(0)\delta^{(2)}(k_{1}+k_{2})\mathcal{G}_{F}(k_{1})\int_{k}\int_{l}\mathcal{G}_{F}^{2}(l)\\
 & +12\delta^{(2)}(k_{1}+k_{2})\mathcal{G}_{F}(k_{1})\mathcal{G}_{F}(k_{2})\int_{l}\mathcal{G}_{F}(l)\Big],
\end{split}
\end{equation}
where $\int_{l}\mathcal{G}_{F}(l)$ is computed earlier from which $\int_{l}\mathcal{G}_{F}^{2}(l)$ can be computed by taking derivative \textit{w.r.t} the mass. Similarly, $\mathcal{O}(\lambda\sqrt{\lambda})$ is the following
\begin{equation}
\begin{split}
24mi\lambda & \sqrt{\lambda}\int_{l_{1}}\int_{l_{2}}\int_{l_{3}}\frac{1}{l_{1}^{1}+l_{2}^{1}+l_{3}^{1}}\Big[(\delta^{(2)}(l_{3}+l_{1})\delta^{(2)}(l_{2}+k_{2})+\delta^{(2)}(l_{1}+l_{2})\delta^{(2)}(l_{3}+k_{2}))\mathcal{G}_{F}(l_{2})\mathcal{G}_{F}(l_{3})\\
 & +\delta^{(2)}(l_{3}+l_{2})\delta^{(2)}(l_{1}+k_{2})\mathcal{G}_{F}(l_{1})\mathcal{G}_{F}(l_{3})+\delta^{(2)}(l_{3}+k_{2})\delta^{(2)}(l_{2}+k_{1})\mathcal{G}_{F}(l_{2})\mathcal{G}_{F}(l_{3})\frac{\mathcal{G}_{F}(l_{1})}{l_{1}^{1}}\\
 & +\delta^{(2)}(l_{3}+k_{2})\delta^{(2)}(l_{1}+k_{1})\mathcal{G}_{F}(l_{1})\mathcal{G}_{F}(l_{3})\frac{\mathcal{G}_{F}(l_{2})}{l_{2}^{1}}+\delta^{(2)}(l_{2}+k_{2})\delta^{(2)}(l_{3}+k_{1})\mathcal{G}_{F}(l_{2})\mathcal{G}_{F}(l_{3})\frac{\mathcal{G}_{F}(l_{1})}{l_{1}^{1}}\\
 & +\delta^{(2)}(l_{2}+k_{2})\delta^{(2)}(l_{1}+k_{1})\mathcal{G}_{F}(l_{2})\mathcal{G}_{F}(l_{1})\frac{\mathcal{G}_{F}(l_{3})}{l_{3}^{1}}+\delta^{(2)}(l_{1}+k_{2})\delta^{(2)}(l_{3}+k_{1})\mathcal{G}_{F}(l_{1})\mathcal{G}_{F}(l_{3})\frac{\mathcal{G}_{F}(l_{2})}{l_{2}^{1}}\\
 & +\delta^{(2)}(l_{1}+k_{2})\delta^{(2)}(l_{2}+k_{1})\mathcal{G}_{F}(l_{2})\mathcal{G}_{F}(l_{1})\frac{\mathcal{G}_{F}(l_{3})}{l_{3}^{1}}-\delta^{(2)}(l_{3}+l_{1})\delta^{(2)}(l_{2}+k_{1})\mathcal{G}_{F}(l_{2})\mathcal{G}_{F}(l_{3})\frac{\mathcal{G}_{F}(k_{2})}{k_{2}^{1}}\\
 & -\delta^{(2)}(l_{3}+l_{1})\delta^{(2)}(k_{2}+k_{1})\mathcal{G}_{F}(l_{3})\mathcal{G}_{F}(k_{2})\frac{\mathcal{G}_{F}(l_{2})}{l_{2}^{1}}-\delta^{(2)}(l_{1}+l_{2})\delta^{(2)}(l_{3}+k_{1})\mathcal{G}_{F}(l_{2})\mathcal{G}_{F}(l_{3})\frac{\mathcal{G}_{F}(k_{2})}{k_{2}^{1}}\\
 & -\delta^{(2)}(l_{1}+l_{2})\delta^{(2)}(k_{2}+k_{1})\mathcal{G}_{F}(l_{2})\mathcal{G}_{F}(k_{2})\frac{\mathcal{G}_{F}(l_{3})}{l_{3}^{1}}-\delta^{(2)}(l_{2}+l_{3})\delta^{(2)}(k_{2}+k_{1})\mathcal{G}_{F}(l_{3})\mathcal{G}_{F}(k_{2})\frac{\mathcal{G}_{F}(l_{1})}{l_{1}^{1}}\\
 & -\delta^{(2)}(l_{2}+l_{3})\delta^{(2)}(l_{1}+k_{1})\mathcal{G}_{F}(l_{1})\mathcal{G}_{F}(l_{3})\frac{\mathcal{G}_{F}(k_{2})}{k_{2}^{1}}\Big].
\end{split}
\end{equation}
There is another $\mathcal{O}(\lambda\sqrt{\lambda})$ term coming from the following
\begin{equation}
\begin{split}
-2im\lambda & \sqrt{\lambda}\int_{l_{1}}\int_{l_{2}}\int_{l_{3}}\int_{l_{4}}\int_{q_{1}}\int_{q_{2}}\int_{q_{3}}\delta^{(2)}(l_{1}+l_{2}+l_{3}+l_{4})\frac{1}{q_{1}^{1}+q_{2}^{1}+q_{3}^{1}}\frac{\delta}{i\delta j(-k_{1})}\frac{\delta}{i\delta j(-k_{2})}\\
 & \times\frac{\delta}{i\delta j(-l_{1})}\frac{\delta}{i\delta j(-l_{2})}\frac{\delta}{i\delta j(-l_{3})}\frac{\delta}{i\delta j(-l_{4})}\frac{\delta}{i\delta j(-q_{1})}\frac{\delta}{i\delta j(-q_{2})}\frac{\delta}{i\delta j(-q_{3})} \ e^{i\int_{k}j(-k)\mathcal{G}_{F}(k)j(k)}\Bigg|_{j=0}, 
\end{split}
\end{equation}
however, the above term turns out to be $0$ since we have odd number of functional derivatives acting on the exponential of the quadratic form, and then taking the source to be vanishing identically.

\subsection{Three-point function}
Similar to the earlier definition, the three-point function can be expressed in the  following way in the momentum space
\begin{equation}
\begin{split}
\langle\phi_{<}(k_{1}) & \phi_{<}(k_{2})\phi_{<}(k_{3})\rangle =\frac{1}{\mathcal{Z}_{\text{free}}[0]}\frac{\delta}{i\delta j(-k_{1})}\frac{\delta}{i\delta j(-k_{2})}\frac{\delta}{i\delta j(-k_{3})}\mathcal{Z}[j]\Big|_{j=0}\\
 & =\frac{\delta}{i\delta j(-k_{1})}\frac{\delta}{i\delta j(-k_{2})}\frac{\delta}{i\delta j(-k_{3})}e^{i\lambda\int_{l_{1},l_{2},l_{3},l_{4}}\delta^{(2)}(l_{1}+\ldots+l_{4})\frac{\delta}{i\delta j(-l_{1})}\frac{\delta}{i\delta j(-l_{2})}\frac{\delta}{i\delta j(-l_{3})}\frac{\delta}{i\delta j(-l_{4})}}\\
\times & e^{-2m\sqrt{\lambda}\int_{l_{1},l_{2},l_{3}}\frac{1}{l_{1}^{1}+l_{2}^{1}+l_{3}^{1}}\frac{\delta}{i\delta j(-l_{1})}\frac{\delta}{i\delta j(-l_{2})}\frac{\delta}{i\delta j(-l_{3})}}e^{\int_{k}\left(ij(-k)\mathcal{G}_{F}(k)j(k)+j(-k)\mathcal{G}_{F}(k)\frac{6\lambda}{k^{1}}-j(k)\mathcal{G}_{F}(k)\frac{6\lambda}{k^{1}}\right)}\Big|_{j=0}.
\end{split}
\end{equation}
Note that the $\mathcal{O}(\lambda)$ contribution vanishes since the odd number of functional derivatives acting on $e^{i\int_{k}j(-k)\mathcal{G}_{F}(k)j(k)}$, and then we put $j=0$. This is vanishing according to Wick's theorem. The first non-zero contribution comes from the following
\begin{equation}
\begin{split}
-16mi\sqrt{\lambda} & \int_{l_{1},l_{2},l_{3}}\frac{1}{l_{1}^{1}+l_{2}^{1}+l_{3}^{1}}\Big[\mathcal{G}_{F}(l_{3})\mathcal{G}_{F}(l_{2})\mathcal{G}_{F}(k_{1})\delta^{(2)}(l_{1}+l_{3})\delta^{(2)}(l_{2}+k_{3})\delta^{(2)}(k_{1}+k_{2})\\
+\mathcal{G}_{F}(l_{3})\mathcal{G}_{F}(l_{2}) & \mathcal{G}_{F}(k_{1})\delta^{(2)}(l_{1}+l_{2})\delta^{(2)}(l_{3}+k_{3})\delta^{(2)}(k_{1}+k_{2})+\mathcal{G}_{F}(l_{3})\mathcal{G}_{F}(l_{1})\mathcal{G}_{F}(k_{1})\delta^{(2)}(l_{2}+l_{3})\delta^{(2)}(l_{1}+k_{3})\delta^{(2)}(k_{1}+k_{2})\\
+\mathcal{G}_{F}(l_{3})\mathcal{G}_{F}(l_{2}) & \mathcal{G}_{F}(l_{1})\delta^{(2)}(l_{3}+k_{3})[\delta^{(2)}(l_{2}+k_{2})\delta^{(2)}(k_{1}+l_{1})+\delta^{(2)}(l_{1}+k_{2})\delta^{(2)}(k_{1}+l_{2})]\\
+\mathcal{G}_{F}(l_{3})\mathcal{G}_{F}(l_{2}) & \mathcal{G}_{F}(l_{1})\delta^{(2)}(l_{2}+k_{3})[\delta^{(2)}(l_{3}+k_{2})\delta^{(2)}(k_{1}+l_{1})+\delta^{(2)}(l_{1}+k_{2})\delta^{(2)}(k_{1}+l_{3})]\\
+\mathcal{G}_{F}(l_{3})\mathcal{G}_{F}(l_{2}) & \mathcal{G}_{F}(l_{1})\delta^{(2)}(l_{1}+k_{3})[\delta^{(2)}(l_{2}+k_{2})\delta^{(2)}(k_{1}+l_{3})+\delta^{(2)}(l_{3}+k_{2})\delta^{(2)}(k_{1}+l_{2})]\\
+\mathcal{G}_{F}(l_{3})\mathcal{G}_{F}(l_{2}) & \mathcal{G}_{F}(k_{3})\delta^{(2)}(l_{1}+l_{3})[\delta^{(2)}(l_{2}+k_{2})\delta^{(2)}(k_{3}+k_{1})+\delta^{(2)}(l_{2}+k_{1})\delta^{(2)}(k_{3}+k_{2})]\\
+\mathcal{G}_{F}(l_{3})\mathcal{G}_{F}(l_{2}) & \mathcal{G}_{F}(k_{3})\delta^{(2)}(l_{1}+l_{2})[\delta^{(2)}(k_{3}+k_{2})\delta^{(2)}(k_{1}+l_{3})+\delta^{(2)}(l_{3}+k_{2})\delta^{(2)}(k_{1}+k_{3})]\\
+\mathcal{G}_{F}(l_{3})\mathcal{G}_{F}(l_{1}) & \mathcal{G}_{F}(k_{3})\delta^{(2)}(l_{1}+l_{3})[\delta^{(2)}(k_{3}+k_{2})\delta^{(2)}(k_{1}+l_{1})+\delta^{(2)}(l_{1}+k_{2})\delta^{(2)}(k_{1}+k_{3})]\Big].
\end{split}
\end{equation} 
This can be simplified as follows
\begin{equation}
\begin{split}
16mi\sqrt{\lambda} & \Bigg[3\int_{l}\mathcal{G}_{F}(l)\mathcal{G}_{F}(k_{1})
\frac{\mathcal{G}_{F}(k_{3})}{k_{3}^{1}}\delta^{(2)}(k_{1}+k_{2})+6\frac{1}{k_{1}^{1}+k_{2}^{1}+k_{3}^{1}}\mathcal{G}_{F}(k_{1})\mathcal{G}_{F}(k_{2})\mathcal{G}_{F}(k_{3})\\
 & +2\int_{l}\mathcal{G}_{F}(l)\delta^{(2)}(k_{3}+k_{1})\frac{\mathcal{G}_{F}(k_{2})}{k_{2}^{1}}\mathcal{G}_{F}(k_{3})+2\int_{l}\mathcal{G}_{F}(l)\delta^{(2)}(k_{3}+k_{2})\frac{\mathcal{G}_{F}(k_{1})}{k_{1}^{1}}\mathcal{G}_{F}(k_{3})\\
 & -\int_{l}\frac{1}{l^{1}}\delta^{(2)}(k_{3}+k_{2})\mathcal{G}_{F}^{2}(k_{1})\mathcal{G}_{F}(k_{3})-\int_{l}\frac{1}{l^{1}}\delta^{(2)}(k_{3}+k_{1})\mathcal{G}_{F}^{2}(k_{2})\mathcal{G}_{F}(k_{3})\Bigg],
\end{split}
\end{equation}
where $\int_{l}\frac{1}{l^{1}}=\delta(0)\ln\frac{\Lambda_{0}}{\Lambda_{IR}}$ and all the $\delta(0)$ can later be replaced by volume of the lattice space size. Further, it is important here to note that the above term violates the conservation of 4-momentum among the three external momentum vectors, which arises from the low-energy effective action itself. There is another contribution of the $\mathcal{O}(\lambda\sqrt{\lambda})$ which is the  following 
\begin{equation}
\begin{split}
\frac{\delta}{i\delta j(-k_{1})} & \frac{\delta}{i\delta j(-k_{2})}\frac{\delta}{i\delta j(-k_{3})}\times(-4mi\lambda\sqrt{\lambda})\int_{l_{1},l_{2},l_{3}}\frac{1}{l_{1}^{2}+l_{2}^{1}+l_{3}^{1}}\frac{\delta}{i\delta j(-l_{1})}\frac{\delta}{i\delta j(-l_{2})}\frac{\delta}{i\delta j(-l_{3})}\\
\times\int_{q_{1},q_{2},q_{3},q_{4}}\delta^{(2)} & (q_{1}+\ldots+q_{4})\frac{\delta}{i\delta j(-q_{1})}\frac{\delta}{i\delta j(-q_{2})}\frac{\delta}{i\delta j(-q_{3})} \frac{\delta}{i\delta j(-q_{4})}e^{i\int_{k}j(-k)\mathcal{G}_{F}(k)j(k)}\Bigg|_{j=0}.
\end{split}
\end{equation}
This is difficult to compute analytically since there are ten functional derivatives. It is important here to emphasize here that the coefficients, depending on the $\Lambda_{0}$ and $\Lambda_{\text{IR}}$ in the above-computed correlation functions can be eliminated using the other correlations functions between the field variables located at different points. Hence, all the physical observables do not depend on these two scales $\Lambda_{0}$ and $\Lambda_{\text{IR}}$. For example, up to $\mathcal{O}(\sqrt{\lambda})$ the expression in (\ref{12.1}) can be expressed as
\begin{equation}
\frac{(\langle\phi_{<}(k)\rangle-\langle\phi_{<}(k)\rangle^{(0)})k^{1}}{28m\sqrt{\lambda}\mathcal{G}_{F}(k)}=\frac{(\langle\phi_{<}(l)\rangle-\langle\phi_{<}(l)\rangle^{(0)})l^{1}}{28m\sqrt{\lambda}\mathcal{G}_{F}(l)}.
\end{equation}

\section{Discussion}
In the first part of this article, we did a review of the coherent state description of the solitons. Here, an important relation between the topological charge and the residue of the function $N_{k}$ about the pole of zero momentum mode is shown. The zero momentum modes are also known as soft particles, playing an important role in capturing the long-range behaviour of the system.

The low-energy effective field theoretical description of excitations about kink-solitons is also established in the present article using Wilson's method \cite{sonoda2006wilson, bonini1997wilson, kim1998wilson} of integrating out the high-energy modes. In order to understand the long-range behaviour of the solitons existing in many physical systems as mentioned in \cite{Puga_1982, 2012arXiv1208.4914K, doi:10.1080/00150190701735588}, we need a low-energy description of the UV theory described in a lattice. In this effective field theory, we have found out new terms which are namely, the source term, $\phi^{3}$ like term, and also the modified quadratic term. The presence of these terms is showing new features in terms of the pole structure of the two-point function. We have also found terms violating the 4-momentum conservation, which is expected because of the non-trivial and non-linear configuration of kink-soliton about which we studied the low-energy excitations. Further, the one-point, two-point, and three-point correlation functions up to certain integral forms in momentum space are also computed. These integral forms contain certain disconnected terms. It is also shown that the one-point function in this theory is non-zero, a new feature that can be used in order to identify the presence of soliton-like classical configuration although this vanishes in a large time limit because of the presence of the complex pole in the momentum space of Green-function.

Although the normal modes about static kink configurations are chosen to be also static, the integral over time coordinate only gives rise to a time-scale in which the low-energy processes take place. However, it does not just arise as a multiplication constant in the low-energy effective action of excitations. Propagating degrees of excitations in time are considered without which the degrees of freedom in terms of their energies cannot be separated. 

Further, we have also shown that for arbitrary $\{k_{i}\}$s the one-, two-, and three-point functions are non-zero even when $k_{1}+k_{2}+k_{3}\neq0$ which is nothing but the violation of momentum conservation. This is similar to the Umklapp process \cite{byrne2002role, zulicke2000umklapp} happens in the solids due to the phonons. This feature can also be used in detecting soliton-like classical configurations in many-body systems.

Since soliton configurations cannot be generated from a trivial vacuum configuration in a perturbative manner, hence, the excitations about these non-linear configurations cannot also be generated from the trivial vacuum excitations. This can be seen both from their topological nature of quanta, shown earlier and the non-trivial correlation functions computed up to a certain mathematical form. Recently, the existence of these solitons is shown in non-linear lattices \cite{kartashov2011solitons, kartashov2009vector, kartashov2009two}. This naturally raises the question of studying elementary excitations about these solitonic configurations and their properties like dispersions, correlations, scattering, etc.

\section{Acknowledgement}
SM would like to thank IISER Kolkata for supporting this work through a doctoral fellowship.

\bibliographystyle{unsrt}
\bibliography{solitons}

\end{document}